\documentclass[pra,amsmath,amssymb,floatfix]{revtex4}
\usepackage{amsmath}
\usepackage{graphicx}
\usepackage{epsfig}

\usepackage{bm}

\newcommand{\be}{\begin{equation}}
\newcommand{\ee}{\end{equation}}

\newcommand{\beq}{\begin{eqnarray}}
\newcommand{\eeq}{\end{eqnarray}}

\def\eq#1{(\ref{#1})}
\def\H1{\widehat{H}_1}

\begin{document}

\title{Scattering of massless particles in one-dimensional chiral channel}

\author{Mikhail Pletyukhov$^1$ and  Vladimir Gritsev$^2$}
\affiliation{$^1$Institute for Theory of Statistical Physics, 
RWTH Aachen, 52056 Aachen, 
Germany and JARA -- Fundamentals of Future Information Technology
\\ $^2$Physics Department, University of Fribourg, Chemin du
Musee 3, 1700 Fribourg, Switzerland}

\date{\today}                                           

\begin{abstract}
We present a general formalism describing a propagation of an arbitrary multiparticle wave packet in a one-dimensional multimode chiral channel coupled to an ensemble of emitters which are distributed at arbitrary positions. The formalism is based on a direct and exact resummation of diagrammatic series for the multiparticle scattering matrix. It is complimentary to the Bethe Ansatz and to approaches based on equations of motion, and it reveals a simple and transparent structure of scattering states. In particular, we demonstrate how this formalism works on various examples, including scattering of one-  and two-photon states off two- and three-level emitters, off an array of emitters as well as scattering of coherent light. We argue that this formalism can be constructively used for  study of scattering of an arbitrary initial photonic state off emitters with arbitrary degree of complexity.
\end{abstract}

\maketitle

\section{Introduction}
Recent advance in fabricating quasi one-dimensional (1D) nanostructures has stimulated and motivated extensive theoretical works on propagating photons in reduced dimensions. One of the goals of this research is to achieve considerable photonic nonlinearities for a single photon pass through the structure which would become a fertile platform for variety of applications in future technology and quantum information \cite{1photon-nonlin}. The only way nonlinear effects may show up in the photonic component of scattering is through interaction with ensemble of emitters. This can be achieved by inserting emitters into the 1D channel or by side coupling to evanescent modes of the channel. The scattering probability of an individual quantum particle of wavelength $\lambda$ in the 1D channel is controlled by the ratio $\lambda^{2}/A$ where $A$ is the effective scattering area of the channel. Therefore reducing the effective scattering area makes nonlinear effects feasible in the range of wavelengths close to $\lambda$. Several physical realizations of such kind of scattering have been recently suggested: tapered optical fibers with atomic ensemble coupled to the evanescent field (currently tapered region can have optical sub wavelength diameters down to 50 nm) \cite{fibers}; hollow optical fibers, "stuffed" with cold atoms \cite{hollow}, photonic crystals \cite{phot-crystal}. Besides conventional atomic ensembles, various artificial emitters and channels can be used as well. Thus, superconducting q-bits coupled to transmission lines have been engineered  to demonstrate two- and three-level emitters (see \cite{supercond} for extensive recent review). Moreover, photons can be replaced by surface plasmons coupled to quantum dots \cite{Akimov}. 

Having in mind these and other physical realizations we consider the model Hamiltonian which describes interaction of a massless multimode quantized bosonic field propagating unidirectionally in 1D geometry with an array of emitters. The coupling between bosons and emitters has the standard dipole-like form. The emitters are assumed to have a multilevel structure (in particular, in this paper we focus on two- and three-level structures, although our formalism allows for a straightforward generalization to more complicated cases as well).  We do not impose any specific details on the spatial distribution of emitters in the 1D channel. However, we distinguish between the case of Dicke arrangement (when all emitters are placed in a space region of the size smaller than a typical wavelength) and the case of remotely distributed emitters. The problem we focus on is of the scattering type: an initial state consists of a direct product of the photonic state and the ground state of an atomic ensemble. Typically, this state is an eigenstate of the free (noninteracting) Hamiltonian. An interaction between light and matter is assumed to be switched on adiabatically in the infinite past and switched off adiabatically in the infinite future. The adiabaticity parameter controls the so-called on-shell condition which plays an important role in the scattering matrix approach (more precisely, the adiabaticity parameter defines an uncertainty in the energy conservation condition). The outgoing state is obtained from the incoming state by an application of a multiparticle scattering matrix. The main goal is to evaluate the scattering matrix, and this is the primary goal of our paper. 

We provide a generic solution of the problem (see Eq.~\eq{eq:main-formula} below) under the assumptions specified above. It can be further extended to the case of multiple emitters. The  coupling constant, detuning, and the level structure may vary from emitter to emitter. Arbitrary initial condition can also be studied, however we specifically focus here on the Fock and coherent states. We found that the multiparticle scattering matrix generically contains many-body bound states in the photonic sector, while in the atomic sector it has a projector-like structure. The result of the scattering consists in preparing the ensemble of scatterers in a specific linear combination of levels, which is nothing else but the dark state, i.e. the state which does not emit. 

Previous studies in eighties \cite{RuYu},\cite{Yu-preprint},\cite{Y2}, nineties \cite{LeClair}, and 00's \cite{1dscat},\cite{WS},\cite{Roy},\cite{Shi},\cite{busch},\cite{YR} have already clarified a number of theoretical questions related to propagating of photons in 1D geometry. Many of these studies have been based on the property of integrability. Our analysis here extends these studies in various directions: we consider arrays of emitters with  multilevel structures and arbitrary coupling constants, and treat the problem in full generality using the methodology of the scattering formalism. Our approach does not require any apriori knowledge  about system's integrability. Nevertheless, our results agree with those obtained by integrability methods, whenever the latter are available. 

In the next section we first present our general result with outline of its derivation. In the following sections we demonstrate its application on specific examples of light scattering off two- and three- level systems. We consider separately the scattering of Fock states with a well-defined number of photons $N$ and the scattering of coherent light. The latter study serves the two purposes: first, the scattering matrix in the coherent state basis  describes a scattering problem of coherent light itself, and, second, it can be considered as a generating functional for scattering problems in  sectors with a well-defined number of photons.  

\section{General solution\label{sec:general-scat-form}}
\subsection{Main result}
We consider the following Hamiltonian
\beq\label{eq:mainH}
H  = \int d \nu \, \nu \, a^{\dagger} (\nu ) a (\nu ) + \sum_{i}
\epsilon_i P_i +\sum_{j=1}^{M} \int d \nu g_{j}\{ a^{\dagger} (\nu ) S^{-}_{j} + a (\nu)
S^{+}_{j} \} \equiv H_0 + a_{\alpha \nu} S_{\bar{\alpha}},
\eeq
where the spectrum of photons is chiral, that is it contains a single branch $\omega=+\nu$ corresponding to a unidirectional propagation of right-moving photons created by $a^{\dagger} (\nu)$. Diagonal $P_i$ and off-diagonal $S_j^{+} = (S_j^{-})^{\dagger}$ operators act in the Hilbert space of the atomic system, and we use the notations $a^{\dagger} =
a_+$, $a = a_-$, $\alpha=\pm$, $\bar{\alpha}=-\alpha$. Here $M$ is the number of atoms (emitters) which are localized in a small spatial region close to $x=0$.  The precise form of operators $S^{a}$ is determined by the level structure of an emitter. Thus, for the two-level system in the rotating wave approximation (RWA) they read $S^{a}_{j}=\frac{\sigma^{a}_{j}}{2}$, where $\sigma^{a}$ are the Pauli matrices. In case of three-level emitters the form of operators $S^{a}$ depends on a specific type of levels' structure, being either of $\Lambda$, $V$, or $\Sigma$-type, which will be  introduced below.  Moreover, eventual counterrotating terms may be also included in these generic notations (in particular, for the two-level system beyond the RWA we would then have $S^{+}=S^{-}=\sigma^{x}/2$). 

The main assumptions we make by writing up the Hamiltonian (\ref{eq:mainH})  are the following: 
1) a linear, unidirectional nature of the photon's spectrum and an absence of its lower and upper bounds; 
2) the interaction is linear in $a_{\alpha}$; 
3) a coupling constant (absorbed in $S_{\bar{\alpha}}$) is
independent of momentum (energy) $\nu$.

Generically, the scattering matrix is defined as a limit $t\rightarrow-\infty$, $t'\rightarrow+\infty$ of the unitary evolution exponent $U(t,t')={\cal T}\exp(-i\int_{t}^{t'}H(\tau)d\tau)$ and contains different scattering channels. The nontrivial part of scattering is contained in the so-called $T$-matrix, which is related to the scattering matrix $S$ via the expression
\be
S = \hat{1} - 2  \pi i \delta (E_{in}-E_{out}) T (E_{in}),
\ee
$E_{in}$ and $E_{out}$ being the energies of incoming and outgoing states.

Under the assumptions we stated above we show that the $T$-matrix in the
$N$-particle sector equals
\beq\label{eq:main-formula}
T^{(N)} (\omega) = G_0^{-1} G (\hat{a}_{\alpha_1 \nu_1}
S_{\bar{\alpha}_1}) G \ldots G (\hat{a}_{\alpha_{2 N} \nu_{2 N}}
S_{\bar{\alpha}_{2 N}}) G G_0^{-1},
\eeq
where the summation over the set of $\alpha_i = \pm$ and integration over the
set of $\nu_i$ is implicitly assumed. The hats over the operators $a$ mean that
they can only be contracted with {\it external} operators (that is, operators creating incoming and outgoing states). The operators $\hat{a}$ are effectively
normal-ordered in \eq{eq:main-formula}, but it is still necessary to account for their commutation relations with $a^{\dagger} a$ appearing in the bare $G_0^{-1} = \omega - H_0$ and dressed $G^{-1}$ Green's functions. In turn, the latter amounts 
to
\be
G^{-1} (\omega)= G_0^{-1} (\omega) - \Sigma = \omega - H_0 + i \pi
S_+ S_- .
\ee

We state that if the {\it self-energy} $\Sigma = - i \pi S_+ S_-$ does
not have a zero eigenvalue, then $T (\omega)$ identically vanishes
when being put on shell ($\omega=E_{in}=E_{out}$): since both $E_{in}$ and $E_{out}$ are the eigenstates of $H_0$, then $G_0^{-1} |_{\mathrm{os}} =0$,
while $G |_{\mathrm{os}}$ has finite eigenvalues in all atomic states. 
To better clarify this important statement we note that the combination $G_{0}^{-1}G$ is nothing else, but the projector onto the atomic states with zero broadening (dark states). Therefore the first task is to evaluate this building block of \eq{eq:main-formula}. 

We also note that from the diagrammatic point of view, the class of models we study here does not allow for diagrams with intersection and overlapping of photonic lines, what follows from the spectrum linearity and the causality imposed by the absence of backscattering for propagating chiral modes. It is naturally possible to extend a diagrammatic approach beyond this class, but then the non-crossing approximation no longer provides an exact solution, and one should expect effects associated with vertex corrections to the vertices $V$.

\subsection{Derivation of the main result}
Below we derive our main expression (\ref{eq:main-formula}). First we remind  several basic facts about  the scattering matrix approach illustrating them on a simple example of scattering off the two-level system. After that, we outline the main steps of our derivation. 
\subsubsection{Scattering problem}
The main goal of the scattering theory is to calculate the scattering matrix
\be
S = {\cal T}\exp \left[ - i \int_{-\infty}^{+\infty} V (t) dt\right] ,
\label{Smatrix-def}
\ee
where the interaction term $V(t)$ is evaluated in the interaction picture. The matrix elements of the scattering matrix $S_{n'n} = \langle n' | S | n \rangle$ are defined in
the states of the non-interacting Hamiltonian, which in our case have the following form
\beq
| n \rangle &=& a^{\dagger}_{k_1} \ldots a^{\dagger}_{k_n} |0 \rangle_b | \sigma \rangle, \quad \varepsilon_{n} = \sum_{i=1}^n k_{i} + \sigma \frac{\Omega}{2}, \\
| n' \rangle &=& a^{\dagger}_{p_1} \ldots a^{\dagger}_{p_{n'}} |0
\rangle_b | \sigma' \rangle, \quad \varepsilon_{n'} =
\sum_{i=1}^{n'} p_{i} + \sigma' \frac{\Omega}{2},
\eeq
where $|0 \rangle_b$ is the photon vacuum state, and $\sigma = \pm$
labels the states of the two-level system, and $\Omega$ is the level splitting.

By definition the $T$-matrix is
\be
T (\omega) = V + V \hat{G} (\omega) V,
\label{Tmatrix-def}
\ee
where the full Green's function is defined as $\hat{G} (\omega) =  (\omega - H + i \eta)^{-1}$.  The parameter $\eta$ controls the adiabaticity of switching the interaction on and off in the far past and far future, respectively. The matrix elements of the $S$-matrix \eq{Smatrix-def} and the $T$-matrix
(\ref{Tmatrix-def}) are related to each other by the the following equation, $
\langle n' |S |n \rangle = \langle n' | n \rangle - 2 \pi i \delta
(\varepsilon_n - \varepsilon_{n'} ) \langle n' | T (\varepsilon_n )
| n \rangle $.
Therefore, it is sufficient to calculate the matrix elements of the
on-shell $T$-matrix  (that is, at $\omega = \varepsilon_n$).

\subsubsection{Calculation of the $T$-matrix}
We expand $\hat{G}$ in (\ref{Tmatrix-def}) in a series of the interaction $V$
\be
\hat{G} (\omega) = G_0 (\omega) + G_0 (\omega ) V G_0 (\omega) + G_0
(\omega ) V G_0 (\omega) V G_0 (\omega) + \ldots ,
\ee
where
\be
G_0 (\omega) = \frac{1}{\omega - H_0 + i \eta} = \frac{P_+}{\omega
-\frac{\Omega}{2} - H_b + i \eta} + \frac{P_-}{\omega +
\frac{\Omega}{2} - H_b + i \eta} .
\ee
is the bare Green's function. The projectors $P_{\pm} = \sigma_{\pm} \sigma_{\mp} = \frac12 (1\pm
\sigma_z)$ map onto the spin states $|\pm \rangle$, respectively.
We note that  $P_{\pm} \sigma_{\mp} = \sigma_{\pm} P_{\pm} = 0 $, and
$P_{\pm} \sigma_{\pm} = \sigma_{\pm} P_{\mp} $, and therefore $T  = V + V G_0 V + V G_0 V G_0 V + \ldots $. 
As $V$ is linear in bosonic operators, we can omit odd powers of
$V$ in this expansion: they do not conserve the number of photons
and will vanish in the calculation of matrix elements. Then
$T = W + W G_0 W + \ldots $,
where $W = V G_0 V$.

Using the properties of $P_{\pm}$, one can show that only diagonal
elements $T_{\pm \pm} = \langle \pm | T | \pm \rangle $ in spin
space are nonzero
\beq
T_{++} (\omega) &=& g^2 a_{\nu'_1} \frac{1}{\omega + \frac{\Omega}{2} -H_b + i \eta} a^{\dagger}_{\nu_1}  \label{tpp} \\
&+& g^4 \left( a_{\nu'_1} \frac{1}{\omega + \frac{\Omega}{2} -H_b +
i \eta} a^{\dagger}_{\nu_1} \right) \frac{1}{\omega -
\frac{\Omega}{2} - H_b + i \eta }
\left( a_{\nu'_2} \frac{1}{\omega + \frac{\Omega}{2} -H_b + i \eta} a^{\dagger}_{\nu_2} \right)+ \ldots , \nonumber \\
T_{--} (\omega) &=& g^2 a_{\nu_1}^{\dagger} \frac{1}{\omega - \frac{\Omega}{2} -H_b + i \eta} a_{\nu'_1}  \label{tmm} \\
&+& g^4 \left( a_{\nu_1}^{\dagger} \frac{1}{\omega -
\frac{\Omega}{2} -H_b + i \eta} a_{\nu'_1} \right) \frac{1}{\omega +
\frac{\Omega}{2} - H_b + i \eta } \left( a_{\nu_2}^{\dagger}
\frac{1}{\omega - \frac{\Omega}{2} -H_b + i \eta} a_{\nu'_2}
\right)+ \ldots . \nonumber
\eeq
Here the integration over the frequencies $\{ \nu_i , \nu'_i \} $ is
implicitly assumed.

Let us put $T_{++}$ and $T_{--}$ partially on-shell introducing
$\omega = \omega_b + \frac{\Omega}{2}$ in \eq{tpp}, and $\omega =
\omega_b - \frac{\Omega}{2}$ in \eq{tmm}, where $\omega_b$ is the
energy of photons, $\omega_b = \sum_i k_i = \sum_i p_i$. Then, we
can rewrite
\beq
T_{++} (\omega_b) &=& g^2 a_{\nu'_1} \frac{1}{\omega_b + \Omega -H_b + i \eta} a^{\dagger}_{\nu_1}  \label{tpps} \\
&+& g^4 \left( a_{\nu'_1} \frac{1}{\omega_b + \Omega -H_b + i \eta}
a^{\dagger}_{\nu_1} \right) \frac{1}{\omega_b - H_b + i \eta }
\left( a_{\nu'_2} \frac{1}{\omega_b + \Omega -H_b + i \eta} a^{\dagger}_{\nu_2} \right)+ \ldots , \nonumber \\
T_{--} (\omega_b ) &=& g^2 a_{\nu_1}^{\dagger} \frac{1}{\omega_b - \Omega -H_b + i \eta} a_{\nu'_1}  \label{tmms} \\
&+& g^4 \left( a_{\nu_1}^{\dagger} \frac{1}{\omega_b - \Omega -H_b +
i \eta} a_{\nu'_1} \right) \frac{1}{\omega_b  - H_b + i \eta }
\left( a_{\nu_2}^{\dagger} \frac{1}{\omega_b - \Omega -H_b + i \eta}
a_{\nu'_2} \right)+ \ldots . \nonumber
\eeq

As we are interested in  evaluation of average values in the bosonic
eigenstates
$\langle a_{p_n} \ldots a_{p_1} T_{\pm \pm } a^{\dagger}_{k_1} \ldots
a^{\dagger}_{k_n} \rangle $,
it is natural to exploit the Wick's theorem, implying
$\langle a_{\nu'} a^{\dagger}_{\nu} \rangle = \delta (\nu- \nu')$ and
$\langle a^{\dagger}_{\nu}  a_{\nu'} \rangle =0 $.
We also note the following intertwining properties of bosonic operators
\beq
a^{\dagger}_{\nu} \frac{1}{\omega-H_b - \nu} = \frac{1}{\omega -H_b}
a^{\dagger}_{\nu}, \quad a_{\nu'} \frac{1}{\omega-H_b + \nu'} =
\frac{1}{\omega -H_b} a_{\nu'}.
\label{comm_prop}
\eeq

An application of the Wick's theorem implies that we have to
contract pairwise the operators $a_{\nu'}$ and $a_{\nu}^{\dagger}$. In order to
do that, it necessary to move $a_{\nu'}$ to the right, commuting it
by means of \eq{comm_prop} with all propagators appearing in between
its initial position and the position of $a_{\nu}^{\dagger}$. There are only the two possibilities for such a contraction: 1)
$a_{\nu}$ is contracted with an {\it adjacent} operator
$a_{\nu'}^{\dagger}$ standing to the right from it; 2) $a_{\nu}$ is
contracted with some {\it external} operator $a^{\dagger}_{k_i}$. In
the diagrammatic representation this means that only diagrams with
non-crossing lines of contraction are allowed. 

The reason for this restriction is the following. If $a_{\nu'}$
is reshuffled with more than one propagator, and after that it is
contracted to some {\it internal} operator, we obtain an integral
over $\nu'$ with more than one pole lying in the same half-plane.
Such an integral identically vanishes. The operator $a_{\nu'}$ can then be 
loosely reshuffled to the right end, which provides the
second possibility. A contraction of the two adjacent operators yields
\be
\langle a_{\nu'} \frac{1}{\omega - H_b + i \eta} a^{\dagger}_{\nu}
\rangle = \langle \int d \nu d \nu' \frac{\delta (\nu -
\nu')}{\omega - H_b - \nu' + i \eta} \rangle = - i \pi .
\ee
This generates the  self-energy insertion $\Sigma$.

Let us denote by $a_{\mu'}$ and $a^{\dagger}_{\mu}$ the operators
which are contracted to the {\it external} operators. In the
$N$-photon sector we have the number $N$ of both species, moreover
for the two-level system
they alternate (due to $\sigma_+^2 = \sigma_-^2 =0$), that is after $a_{\mu'}$ we must have
$a_{\mu}^{\dagger}$, not $a_{\mu''}$. All self-energy
insertions between such pairs can be resummed, and we obtain for
$T_{++}$ and $T_{--}$ the following result in the $N$-particle
sector
\beq
T_{++}^{(N)} (\omega_b ) &=& g^{2 N} (\omega_b - H_b + i \eta)
\frac{1}{\omega_b - H_b + i \pi g^2}
\left( a_{\mu'_1} \frac{1}{\omega_b + \Omega - H_b +i \eta} a_{\mu_1}^\dagger \right) \nonumber \\
& \times & \frac{1}{\omega_b  - H_b + i \pi g^2} \ldots \frac{1}{\omega_b - H_b + i \pi g^2} \nonumber \\
& \times & \left( a_{\mu'_{N}} \frac{1}{\omega_b +\Omega - H_b + i
\eta } a_{\mu_{N}}^\dagger \right)\frac{1}{\omega_b - H_b + i \pi
g^2} (\omega_b - H_b + i \eta),
\eeq
\beq
T_{--}^{(N)} (\omega_b) &=& g^{2N} a_{\mu_1}^{\dagger} \frac{1}{\omega_b - \Omega - H_b + i \pi g^2} \left( a_{\mu'_1} \frac{1}{\omega_b - H_b +i \eta} a_{\mu_2}^\dagger \right) \nonumber \\
& \times & \frac{1}{\omega_b - \Omega - H_b + i \pi g^2} \ldots \frac{1}{\omega_b - \Omega - H_b + i \pi g^2} \nonumber \\
& \times & \left( a_{\mu'_{N-1}} \frac{1}{\omega_b - H_b + i \eta }
a_{\mu_{N}}^\dagger \right) \frac{1}{\omega_b - \Omega - H_b + i \pi
g^2} a_{\mu'_N} .
\eeq
In these relations operators $a$ and $a^{\dagger}$ effectively
commute with each other, and one has only to take carefully into
account permutations of $a (a^{\dagger})$ with the propagators using
\eq{comm_prop}. Note that $T_{++} (\omega_b)$ vanishes, when it is put on-shell,
because of the factors $(\omega_b - H_b +i \eta)$ in the beginning
and in the end of the corresponding expression. 

Let us now consider $T_{--}^{(N)} (\omega_b)$. Moving all propagators to
the left and putting them on-shell, we obtain
\beq
T_{--}^{(N)} &=& g^{2N} \frac{1}{\mu_1 - \alpha} \,\, \frac{1}{\Delta_{11'} + i \eta} \,\, \frac{1}{\Delta_{11'} + \mu_2- \alpha} \,\, \ldots \,\, \frac{1}{\sum_{i=1}^{N-2} \Delta_{ii'} + \mu_{N-1} - \alpha} \nonumber \\
& \times & \frac{1}{\sum_{i=1}^{N-1} \Delta_{ii'} + i \eta } \,\,
\frac{1}{\sum_{i=1}^{N-1} \Delta_{ii'} + \mu_N -\alpha } \,\,
a_{\mu_1}^{\dagger} ... a_{\mu_N}^{\dagger} a_{\mu'_N} \ldots
a_{\mu'_1},
\label{tmmN}
\eeq
where $\Delta_{ii'}  = \mu_i - \mu_{i'}$ and $\alpha = \Omega - i
\pi g^2$.

A direct generalization of this derivation to cases with a more complicated level structure leads to our main equation (\ref{eq:main-formula}).

\section{Examples}
Here we consider several important examples and compute the scattering matrix for the cases of a single emitter coupled to a 1D waveguide within and beyond the RWA. We focus on the two-level and three-level cases. 
\subsection{One emitter in the RWA}

For pedagogical reasons, we repeat here the derivation of \eq{tmmN} from the
general expression \eq{eq:main-formula}.

In the case of a single atom interacting with a 1D field  in the RWA we define
$\epsilon_{\pm} = \pm \frac{\Omega}{2}$, $P_{\pm} = \sigma_{\pm} \sigma_{\mp}$, and 
$S_{\pm} = g \sigma_{\pm} = g \frac{\sigma_x \pm i \sigma_y}{2}$.
We find then $\Sigma = - i \pi S_+ S_- = - i \pi g^2 P_+$ and
$G (\omega) = P_+ (\omega  - \frac{\Omega}{2} - H_b + i \pi
g^2)^{-1} +P_- (\omega + \frac{\Omega}{2} - H_b + i\eta)^{-1} = P_+ G_+
+ P_- G_- $. Therefore, the main building blocks of the equation for the $T$-matrix are given by
\beq
G_0^{-1} G &=&  ( \omega  - \frac{\Omega}{2} - H_b + i \eta ) \frac{P_+}{\omega  - \frac{\Omega}{2} - H_b + i \pi g^2}  +P_- , \\
G G_0^{-1} &=& \frac{P_+}{\omega  - \frac{\Omega}{2} - H_b + i \pi
g^2} ( \omega  - \frac{\Omega}{2} - H_b + i \eta )  +P_- .
\eeq
Putting them on-shell we find
\be
G_0^{-1} G |_{\mathrm{os}} = G G_0^{-1} |_{\mathrm{os}} = P_- ,
\ee
and therefore
\beq
T (\omega) &=& g^{2 N} P_-  (\hat{a}_{+, \nu_1}  \sigma_- ) P_+ G_+ (\hat{a}_{-, \nu_2} \sigma_+) P_- G_- \ldots  P_- G_- (\hat{a}_{+, \nu_{2 N-1}}  \sigma_- ) P_+ G_+ (\hat{a}_{-, \nu_{2N}} \sigma_+) P_- \nonumber \\
&=& P_- g^{2N} \hat{a}^{\dagger}_{\nu_1}  G_+ \hat{a}_{\nu_2} G_-
\ldots  G_- \hat{a}^{\dagger}_{\nu_{2 N-1}}   G_+ \hat{a}_{\nu_{2N}}.
\eeq
Introducing $\omega = \omega_b -\frac{\Omega}{2}$, we can rewrite
$G_{\pm}$ as 
$G_+ = (\omega_b - H_b - \Omega + i \pi g^2)^{-1}$ and $G_- =
(\omega - H_b + i \eta)^{-1} $.

\subsection{$M$ emitters in the RWA (Dicke case)}
In the case of many emitters confined in a small (compared to the typical wavelength) region of space we have
$\epsilon_{m} = m \Omega$,  $P_{m} = |m \rangle \langle m |$, while 
$S_{\pm}  \to   g S_{\pm}$ . The total effective spin is conserved, $S_a S_a = l (l+1)$. If the system is
initially in the ground state, that is $S_z = -M/2$, then it is
sufficient to consider the representation with the largest weight
$l=M/2$, to which the ground state belongs. In this case we find
\be
\Sigma = - i  \pi g^2 S_+ S_- = - i  \pi g^2 (S_a^2 - S_z^2 + S_z) =
- i  \pi g^2 \sum_{m=-l}^l P_m [l (l+1) - m (m-1)],
\ee
and therefore for the dressed Green's function we find
\beq
G (\omega) = \sum_{m=-l}^l \frac{P_m}{\omega- H_b  - m \Omega + i
\pi g^2 (l+m) (l-m+1)}.
\eeq
The main building block is therefore equal to
\be
G_0^{-1} G |_{\mathrm{os}} = G G_0^{-1} |_{\mathrm{os}} = P_{-l} =
|-l \rangle \langle -l |.
\ee
In the one-photon sector we immediately get
\beq
T^{(1)} (\omega) &=& g^2 P_{-l} (a^{\dagger}_{\nu_1} S_- )
P_{-l+1} G_{-l+1} (a_{\nu_2} S_+) P_{-l} \nonumber \\
&=& g^2 P_{-l} |\langle -l |S_- | - l+1 \rangle|^2
a^{\dagger}_{\nu_1} \frac{1}{\omega - H_b + (l-1) \Omega + i 2 \pi
g^2 l} a_{\nu_2}.
\eeq
Using the action of the collective spin operators on the states of emitters
\be
S_{\pm}|m \rangle = \sqrt{(l \mp m) (l \pm m +1)} | m \pm 1 \rangle ,
\label{Spm}
\ee
we obtain that 
$|\langle -l |S_- | - l+1 \rangle|^2 = 2 l $.
Introducing $\omega = \omega_b - \Omega l = \omega_b - \Omega M/2$
as well as $\alpha_M = \Omega - i \pi g^2 M$,
and omitting $P_{-l}$, we obtain $T^{(1)} (\omega_b) = g^2 M a^{\dagger}_{\nu_1} (\omega_b - H_b - \alpha_M)^{-1} a_{\nu_2} $.
Putting this expression on-shell results in 
\be
T^{(1)} =g^2 M \frac{1}{\nu_1 - \alpha_M} a^{\dagger}_{\nu_1}
a_{\nu_2} .
\ee
In the two-photon sector
\beq
T^{(2)} (\omega) &=& g^4 P_{-l} (a^{\dagger}_{\nu_1} S_- ) P_{-l+1}
G_{-l+1} (a_{\alpha_2 \nu_2} S_{\bar{\alpha}_2} ) P_{m_2} G_{m_2}
(a_{\alpha_3 \nu_3} S_{\bar{\alpha}_3} ) P_{-l+1} G_{-l+1}
(a_{\nu_4} S_+) P_{-l} \nonumber \\
&=& g^4 P_{-l} |\langle -l |S_- | - l+1 \rangle|^2 \langle -l+1|
S_{\bar{\alpha}_2} | m_2 \rangle \langle m_2 | S_{\bar{\alpha}_3} |
-l+1 \rangle
\nonumber \\
& \times & a^{\dagger}_{\nu_1} \frac{1}{\omega- H_b + (l-1)\Omega  +
i 2 \pi g^2 l} a_{\alpha_2 \nu_2} \frac{1}{\omega- H_b  - m_2 \Omega
+ i  \pi g^2 (l+m_2 ) (l-m_2 +1)}
\nonumber \\
& \times & a_{\alpha_3 \nu_3 } \frac{1}{\omega - H_b + (l-1)\Omega
+ i 2 \pi g^2 l} a_{\nu_4}.
\eeq
Using \eq{Spm} we obtain
\beq
& &  |\langle -l |S_- | - l+1 \rangle|^2 \langle -l+1| S_{\bar{\alpha}_2} | m_2 \rangle \langle m_2 | S_{\bar{\alpha}_3} | -l+1 \rangle \nonumber \\
&=& (2 l) (\delta_{\alpha_2 -} \delta_{m_2, -l} \sqrt{2 l} + \delta_{\alpha_2 +} \delta_{m_2, -l+2} \sqrt{2 (2 l-1)}) (\delta_{\alpha_3 +} \delta_{m_2, -l} \sqrt{2 l} + \delta_{\alpha_3 -} \delta_{m_2, -l+2} \sqrt{2 (2 l-1)}) \nonumber \\
&=& 2 l [ 2 l \delta_{\alpha_2 -} \delta_{\alpha_3 +} \delta_{m_2 ,
-l} + 2 (2 l -1) \delta_{\alpha_2 +} \delta_{\alpha_3 -} \delta_{m_2
, -l+2}]
\eeq
and
\beq
T^{(2)} (\omega ) &=& P_{-l} g^4  4 l^2 a^{\dagger}_{\nu_1}
\frac{1}{\omega - H_b + (l-1)\Omega  + i 2 \pi g^2 l} a_{ \nu_2}
\frac{1}{\omega - H_b
+ l \Omega + i \eta} \nonumber \\
& \times & a_{ \nu_3 }^{\dagger} \frac{1}{\omega - H_b +  (l-1)\Omega  + i 2 \pi g^2 l}  a_{\nu_4} \nonumber \\
&+& P_{-l} g^4  4 l (2 l- 1) a^{\dagger}_{\nu_1} \frac{1}{\omega -
H_b + (l-1)\Omega  + i 2 \pi g^2 l} a_{ \nu_2}^{\dagger}
\frac{1}{\omega - H_b + (l-2) \Omega + i  \pi g^2 2  (2 l-1)}
\nonumber \\
& \times & a_{ \nu_3 } \frac{1}{\omega - H_b + (l-1)\Omega  + i 2
\pi g^2 l} a_{\nu_4}.
\eeq

Introducing $\alpha_{M-1} = \Omega - i  \pi g^2 (M-1)$
and omitting $P_{-l}$, we arrive at
\beq
T^{(2)} (\omega_b ) &=& g^4  M^2 a^{\dagger}_{\nu_1}
\frac{1}{\omega_b - H_b - \alpha_M} a_{ \nu_2} \frac{1}{\omega_b -
H_b
 + i \eta} a_{ \nu_3 }^{\dagger} \frac{1}{\omega_b - H_b - \alpha_M }  a_{\nu_4} \nonumber \\
&+& g^4  M (M- 1) a^{\dagger}_{\nu_1} \frac{1}{\omega_b - H_b
-\alpha_M} a_{ \nu_2}^{\dagger} \frac{1}{\frac12 (\omega_b - H_b) -
\alpha_{M-1}}
 a_{ \nu_3 } \frac{1}{\omega_b - H_b - \alpha_M} a_{\nu_4}.
\eeq
Putting $T^{(2)}$ on-shell, we obtain (note relabeling $\nu_2
\leftrightarrow \nu_3$ in the first term which is allowed because of the symmetric nature of the photonic wavefunction)
\beq
T^{(2)} &=& g^4  M  \frac{1}{\nu_1 - \alpha_M}  \frac{1}{\nu_1 +\nu_2 - \nu_3  - \alpha_M }  \left[ \frac{M}{\nu_1 - \nu_3 + i \eta}  +
\frac{M-1}{\frac12 (\nu_1 + \nu_2) - \alpha_{M-1}} \right]
a^{\dagger}_{\nu_1} a_{ \nu_2}^{\dagger} a_{ \nu_3 } a_{\nu_4} .
\eeq

In the one-photon sector we have 
\beq
t_{p_1 k_1} = \delta_{p_1 k_1} \langle a_{p_1} T^{(1)}
a_{k_1}^{\dagger} \rangle = \delta_{p_1 k_1} \frac{M g^2}{\nu_1 -
\alpha_M} \langle a_{p_1}a_{\nu_1}^{\dagger} a_{\nu_2}
a_{k_1}^{\dagger} \rangle = \delta_{p_1 k_1} \frac{M g^2}{p_1 -
\alpha_M} ,
\eeq
and therefore the scattering matrix is
\be
S_{p_1 k_1} = \delta_{p_1 k_1} - 2 \pi i t_{p_1 k_1} =  \delta_{p_1
k_1} \left( 1 - 2 \frac{i \pi M g^2}{p_1 - \alpha_M } \right).
\ee

Let us consider the scattering matrix in the two-photon sector. First,
we consider
\be
\delta_{p_1+p_2 ,  k_1 +k_2} \langle a_{p_2} a_{p_1} T^{(1)}
a_{k_1}^{\dagger} a_{k_2}^{\dagger} \rangle = \delta_{p_2 k_2}
t_{p_1 k_1} + \delta_{p_1 k_1} t_{p_2 k_2} + \delta_{p_2 k_1} t_{p_1
k_2} + \delta_{p_1 k_2} t_{p_2 k_1}.
\ee
Second,
\beq
& & \delta_{p_1+p_2 ,  k_1 +k_2} \langle a_{p_2} a_{p_1} T^{(2)} a_{k_1}^{\dagger} a_{k_2}^{\dagger} \rangle = \delta_{p_1+p_2 ,  k_1 +k_2} g^4 M\frac{1}{\nu_1 - \alpha_M} \frac{1}{\nu_4 -\alpha_M} \nonumber \\
& & \times 
 \left[ \frac{M}{\nu_1 - \nu_3 } - i M \pi \delta_{\nu_1 \nu_3}  + \frac{M-1}{\frac12 (\nu_1 + \nu_2) - \alpha_{M-1}} \right] \nonumber \\
& & \times (\delta_{p_1 \nu_1} \delta_{p_2 \nu_2} + \delta_{p_1
\nu_2} \delta_{p_2 \nu_1}) (\delta_{k_1 \nu_4} \delta_{k_2 \nu_3} +
\delta_{k_1 \nu_3} \delta_{k_2 \nu_4}),
\eeq
and therefore
\beq
S_{p_1 p_2 , k_1 k_2} = S_{p_1 k_1} S_{p_2 k_2} + S_{p_2 k_1} S_{p_1
k_2} + i \mathcal{T}^{(2)}_{p_1 p_2 , k_1 k_2} ,
\eeq
where
\beq
\mathcal{T}^{(2)}_{p_1 p_2 , k_1 k_2} &=& - 2 \pi \delta_{p_1+p_2 ,
k_1 +k_2}
g^4 M  \nonumber \\
& & \times \frac{1}{\nu_1 - \alpha_M} \frac{1}{\nu_4 -\alpha_M}
\left[ \frac{M}{\nu_1 - \nu_3 }  + \frac{M-1}{\frac12 (\nu_1 + \nu_2) - \alpha_{M-1}} \right] \nonumber \\
& & \times (\delta_{p_1 \nu_1} \delta_{p_2 \nu_2} + \delta_{p_1
\nu_2} \delta_{p_2 \nu_1}) (\delta_{k_1 \nu_4} \delta_{k_2 \nu_3} +
\delta_{k_1 \nu_3} \delta_{k_2 \nu_4}).
\eeq
Defining
$E= p_1 + p_2 = k_1 + k_2$,  $\Delta = \frac{1}{2}(k_1 - k_2)$, and $\Delta' = \frac{1}{2}(p_1 - p_2)$
we find that 
$k_{1,2} = \frac{E}{2} \pm \Delta $, $p_{1,2} = \frac{E}{2} \pm
\Delta' $.
Therefore the $T$-matrix is
\beq
\mathcal{T}^{(2)}_{p_1 p_2 , k_1 k_2} &=& \frac{8 \pi
\delta_{p_1+p_2 ,  k_1 +k_2}
g^4 M (\frac{E}{2} - \alpha_M)}{[(\frac{E}{2} - \alpha_M)^2 - \Delta^2] [(\frac{E}{2} - \alpha_M)^2 - \Delta'^{2}]} \left\{ M- (M-1) \frac{\frac{E}{2} - \alpha_M}{\frac{E}{2} - \alpha_{M-1}}  \right\} \nonumber \\
&=& \frac{8 \pi \delta_{p_1+p_2 ,  k_1 +k_2} g^4 M (\frac{E}{2} -
\alpha_M)}{[(\frac{E}{2} - \alpha_M)^2 - \Delta^2] [(\frac{E}{2} -
\alpha_M)^2 - \Delta'^{2}]} \frac{\frac{E}{2}}{\frac{E}{2} -
\alpha_{M-1}},
\label{t2finpole}
\eeq
which is in agreement with the result of Ref.~\cite{YR}. The emergence of the poles in \eq{t2finpole} involving more than one individual photon's energy reflects a formation of the photonic bound state.

\subsection{One emitter beyond the RWA}
For the case of an emitter interacting with the field beyond the RWA we have
$S_+ = g \sigma_+ + g' \sigma_-$ and $S_- = g \sigma_- + g' \sigma_+ $. Calculating the self-energy we find
$\Sigma = - i \pi S_+ S_- = -i \pi g^2 P_+ - i \pi  (g')^2 P_-$, and
therefore
$G_0^{-1} G |_{\mathrm{os}} = G G_0^{-1} |_{\mathrm{os}} =0$. This implies that the $T$-matrix identically vanishes.

Thus, we observe that the non-RWA model is not analytically connected to the RWA one. The nonanalyticity is hidden in the noncommutativity of the limits $\eta\rightarrow 0$ (which should be taken first for the non-RWA model) and $g' \to 0$: the corresponding eigenvalue $\eta/(\pi (g')^{2}+\eta)$ of the projector onto the emitter's groundstate is sensitive to the order of these limits, and a smooth crossover between the non-RWA and the RWA models would be only possible for the finite value of $\eta$.

\subsection{Three-level system, $\Lambda$-scheme}
\label{subsec:Lambda}
\begin{figure}[h]
\begin{center}
\includegraphics[width=7cm]{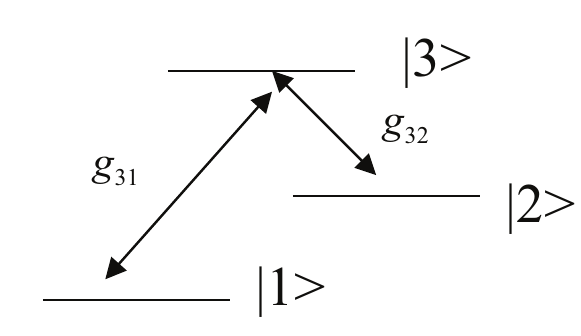}
\caption{Three-level $\Lambda$-scheme}
\label{fig:Lambda}
\end{center}
\end{figure}

For the three-level scheme of 
$\Lambda$-type (see Fig.~\ref{fig:Lambda}) we have $
S_+ = g_{31} |3 \rangle \langle 1| + g_{32} |3 \rangle \langle 2|$, and
$S_- = g_{31} |1 \rangle \langle 3| + g_{32} |2 \rangle \langle 3|$. 
The self-energy is $\Sigma = - i \pi (g_{31}^2 + g_{32}^2) P_3 \equiv - i \pi g^2 P_3$, hence the dressed Green's function acquires the following form
\be
G = \frac{P_1}{\omega - \epsilon_1 - H_b + i \eta} + \frac{P_2}{\omega -
\epsilon_2 - H_b + i \eta} + \frac{P_3}{\omega - \epsilon_3 - H_b + i \pi g^2} =
P_i G_i.
\ee

Therefore the building block is 
\be
G_0^{-1} G |_{\mathrm{os}} = G G_0^{-1} |_{\mathrm{os}} =P_1 +P_2.
\ee
The $T$-matrix in the single-photon sector is
\beq
T^{(1)} (\omega) &=& (P_1 +P_2)  (\hat{a}_{\nu_1}^{\dagger} S_-) P_3  G_3 (\hat{a}_{\nu_2} S_+) (P_1 +P_2) \nonumber \\
&=& \left\{ g_{31}^2 P_1 + g_{32}^2 P_2 + g_{31} g_{32} ( |2 \rangle
\langle 1| + | 1 \rangle \langle 2 | ) \right\}
\hat{a}_{\nu_1}^{\dagger} \frac{1}{\omega - \epsilon_3 - H_b + i \pi g^2}
\hat{a}_{\nu_2},
\eeq
and therefore
\beq
S_{pk} &=& \delta_{pk} - 2 \pi i g_{31}^2 \delta_{pk} \frac{P_1}{p +
\epsilon_1 - \epsilon_3 + i \pi g^2} - 2 \pi i g_{32}^2 \delta_{pk} 
\frac{P_2}{p + \epsilon_2 - \epsilon_3 + i \pi g^2}
\nonumber \\
& &  - 2 \pi i g_{31} g_{32} \left[ \delta_{k+\epsilon_1, p+\epsilon_2} 
\frac{|2 \rangle \langle 1|}{p+ \epsilon_2 - \epsilon_3 + i \pi g^2} 
+ \delta_{p+\epsilon_1, k+\epsilon_2} \frac{|1 \rangle \langle 2|}{p+\epsilon_1 - \epsilon_3 + i \pi g^2} \right].
\eeq

In the two-photon sector of scattering we obtain (cf. Ref.~\cite{Roy})
\beq
T^{(2)} (\omega)  &=& (P_1 + P_2) (\hat{a}_{\nu_1}^{\dagger} S_-)
P_3 G_3
(\hat{a}_{\nu_2} S_+) (P_1 G_1 + P_2 G_2)(\hat{a}_{\nu_3}^{\dagger} S_-)  P_3 G_3 (\hat{a}_{\nu_4} S_+) (P_1 +P_2) \nonumber \\
&=& \left\{ g_{31}^2 P_1 + g_{32}^2 P_2 + g_{31} g_{32} ( |2 \rangle
\langle 1| + | 1 \rangle \langle 2 | ) \right\}
\hat{a}_{\nu_1}^{\dagger} G_3 \hat{a}_{\nu_2} (g_{31}^2 G_1 +
g_{32}^2 G_2) \hat{a}_{\nu_3}^{\dagger} G_3 \hat{a}_{\nu_4} .
\eeq

\subsection{Three-level system, $V$-scheme}
\label{subsec:V}
\begin{figure}[ht]
\begin{center}
\includegraphics[width=7cm]{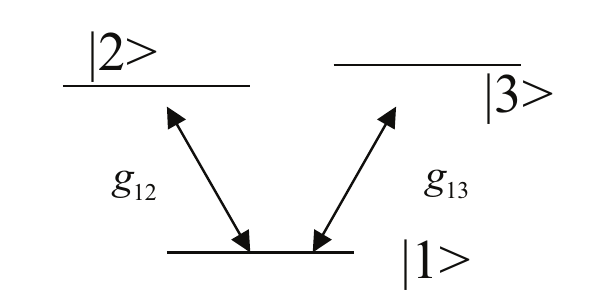}
\caption{Three-level $V$-scheme}
\label{fig:V}
\end{center}
\end{figure}

In the case of three-level $V$-scheme (see Fig.~\ref{fig:V}) the $S_{\pm}$ operators  are
$S_+ = g_{31} |3 \rangle \langle 1| + g_{21} |2 \rangle \langle 1|$, and
$S_- = g_{31} |1 \rangle \langle 3| + g_{21} |1 \rangle \langle 2|$. 
The corresponding self-energy is therefore
$\Sigma = - i \pi  g_{31}^2 P_3 - i \pi g_{21}^2 P_2 - i \pi g_{21}
g_{31} (|2 \rangle \langle 3| + |3 \rangle \langle 2|)$, while the dressed Green's function is
$G = P_1 (\omega - \epsilon_1 - H_b + i \eta)^{-1} + \sum_{a=2,3}
\tilde{P}_a(\omega - H_b - \lambda_a)^{-1}$,
where
\beq
\tilde{P}_2 &=& \xi_{22}^2 P_2 + \xi_{23}^2 P_3 + \xi_{22} \xi_{23} |2 \rangle \langle 3| + \xi_{23} \xi_{22} |3 \rangle \langle 2|, \nonumber\\
\tilde{P}_3 &=& \xi_{33}^2 P_3 + \xi_{32}^2 P_2 + \xi_{32} \xi_{33}
|2 \rangle \langle 3| + \xi_{33} \xi_{32} |3 \rangle \langle 2|,\nonumber\\
\xi_{22} &=& \xi_{33} = \cos \frac{\phi}{2} , \quad \xi_{23} = - \xi_{32} = \sin \frac{\phi}{2}, \nonumber\\
\frac{\lambda_2 - \lambda_3}{2} \cos \phi &=& \frac{\epsilon_2- \epsilon_3}{2} 
- i \pi \frac{g_{21}^2 -g_{31}^2}{2}, \quad \frac{\lambda_2 -
\lambda_3}{2} \sin \phi =
- i \pi g_{21} g_{31} ,\nonumber \\
\frac{\lambda_2 + \lambda_3}{2} &=& \frac{\epsilon_2+ \epsilon_3}{2} - i \pi \frac{g_{21}^2 +g_{31}^2}{2}, \quad \lambda_2 \lambda_3 = (\epsilon_2 - i \pi g_{21}^2) (\epsilon_3 - i \pi g_{31}^2) +\pi^2 g_{21}^2 g_{31}^2 , \nonumber\\
(\lambda - \lambda_2 ) (\lambda - \lambda_3) &=& (\lambda - \epsilon_2 +
i \pi g_{21}^2)(\lambda - \epsilon_3 + i \pi g_{31}^2) + \pi^2 g_{21}^2
g_{31}^2 .
\eeq

Using these relations we observe that our basic building block $G_0^{-1} G |_{\mathrm{os}} = G G_0^{-1} |_{\mathrm{os}} = P_1$,
and therefore
\beq
T^{(1)} (\omega ) &=& P_1 ( \hat{a}^{\dagger}_{\nu_1} S_-)
\tilde{P}_a G_a
(\hat{a}_{\nu_2} S_+)       P_1  \nonumber \\
&=& P_1 \hat{a}^{\dagger}_{\nu_1} \left[
\frac{ \frac{1+\cos \phi}{2} g_{21}^2 + \frac{1 -\cos \phi}{2} g_{31}^2 + g_{21} g_{31} \sin \phi  }{\omega - H_b - \lambda_2} \right. \nonumber \\
& & \left. \qquad + \frac{\frac{1-\cos \phi}{2} g_{21}^2 +
\frac{1+\cos \phi }{2} g_{31}^2 - g_{21} g_{31} \sin \phi }{\omega -
H_b - \lambda_3} \right] \hat{a}_{\nu_2} .
\eeq
Combining the terms in square brackets, we obtain for the single-particle 
scattering matrix
\beq
S_{pk} &=& \delta_{pk} \left[ 1 - 2  \frac{i \pi g_{21}^2 (p+\epsilon_1 - \epsilon_3) + i \pi g_{31}^2 (p+ \epsilon_1 - \epsilon_2)}{(p + \epsilon_1 - \epsilon_2 + i \pi g_{21}^2) (p +\epsilon_1  - \epsilon_3 + i \pi g_{31}^2) +\pi^2 g_{21}^2 g_{31}^2 } \right] \nonumber \\
&=&  \delta_{pk} \frac{(p + \epsilon_1 - \epsilon_2 - i \pi g_{21}^2) (p +\epsilon_1  -\epsilon_3 - i \pi g_{31}^2) +\pi^2 g_{21}^2 g_{31}^2}{(p + \epsilon_1 - \epsilon_2 + i \pi g_{21}^2) (p +\epsilon_1  - \epsilon_3 + i \pi g_{31}^2) +\pi^2 g_{21}^2 g_{31}^2 } .
\eeq
\subsection{Three-level system, $\Sigma$-scheme}
\label{subsec:Sigma}
\begin{figure}[h]
\begin{center}
\includegraphics[width=7cm]{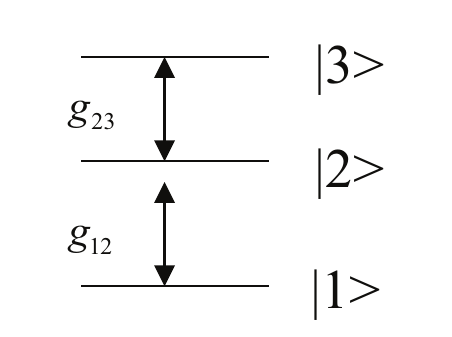}
\caption{Three-level $\Sigma$-scheme}
\label{fig:Sigma}
\end{center}
\end{figure}

In the case of $\Sigma$-scheme (see Fig.~\ref{fig:Sigma}) the role of $S_{\pm}$ operators is played by 
$S_+ = g_{32} |3 \rangle \langle 2| + g_{21} |2 \rangle \langle 1| $, and $S_- = g_{32} |2 \rangle \langle 3| + g_{21} |1 \rangle \langle 2|$.
The corresponding self-energy is
$\Sigma = - i \pi g_{32}^2 P_3 - i \pi g_{21}^2 P_2 $ and therefore
$G = P_1 (\omega - \epsilon_1 - H_b + i \eta)^{-1} + P_2 (\omega -
\epsilon_2 - H_b + i \pi g_{21}^2)^{-1} + P_3 (\omega - \epsilon_3 - H_b + i \pi
g_{32}^2)^{-1} = P_i G_i$. 

The building block is $
G_0^{-1} G |_{\mathrm{os}} = G G_0^{-1} |_{\mathrm{os}} =P_1$, and therefore
\be
T^{(1)} (\omega ) = P_1   (\hat{a}_{\nu_1}^{\dagger} S_- )  P_2 G_2
(\hat{a}_{\nu_2}  S_+ )   P_1 = P_1 g_{21}^2
\hat{a}_{\nu_1}^{\dagger} \frac{1}{\omega - \epsilon_2 + i \pi g_{21}^2 }
\hat{a}_{\nu_2},
\ee
while
\be
S_{pk} = \delta_{pk} \left[ 1 - 2 \pi i \frac{g_{21}^2}{p + \epsilon_1 -
\epsilon_2 + i \pi g_{21}^2}\right] = \delta_{pk} \frac{p + \epsilon_1 - \epsilon_2 -i \pi g_{21}^2}{p + \epsilon_1 - \epsilon_2 + i \pi g_{21}^2} .
\ee
We note that for $\Sigma$-scheme the one-photon scattering is the same as in the case of the two-level model, the third level being inefficient.

The results for the one-photon scattering of subsections D, E, and F agree with the ones derived in Ref.~\cite{WS}.

\section{A model with several emitters \label{sec:diff-pos}}
Here we apply the same formalism to the case of several emitters. We will distinguish the case of distributed emitters  from the case of concentrated system (located at the same point). Moreover we will mostly focus on the case of two-level systems and discuss the cases of single- and two-photon scattering separately. We show that our main result works even when different emitters have different coupling constants. These calculations explicitly prove one of our main statements about an independent character of scattering of unidirectional photons in distributed systems.

\subsection{The case of two atoms}
To be specific we will focus on the following model Hamiltonian with two 
emitters
\beq
H  &=& \int d \nu \, \nu \, a^{\dagger} (\nu ) a (\nu ) + \frac{\Omega_1}{2} \sigma_z^{(1)} + \frac{\Omega_2}{2} \sigma_z^{(2)}  + \int d \nu \{ a^{\dagger} (\nu ) S_- (\nu) + a (\nu) S_+ (\nu) \} \nonumber \\
& \equiv & H_0 + a_{\alpha \nu} S_{\bar{\alpha} \nu},
\eeq
where the following combinations play now the role of spin operators
\beq
S_+ (\nu) &=& g_1 \sigma_+^{(1)} e^{i \nu r_1} + g_2 \sigma_+^{(2)} e^{i \nu r_2} , \\
S_- (\nu) &=& g_1 \sigma_-^{(1)} e^{-i \nu r_1} + g_2 \sigma_-^{(2)}
e^{-i \nu r_2} .
\eeq
It is important to remark that this form of coupling is effectively energy (momentum) dependent, and for $r_1 \neq r_2$ it does not fulfill one of the assumptions we used to derive \eq{eq:main-formula}. Therefore, an extension of \eq{eq:main-formula} is required, which will be done below.

It is convenient to introduce the following projection operators 
$P_{ab} = P_{a}^{(1)} P_b^{(2)}$,  $a,b=\pm$, such that $\sum_{a,b} P_{ab} =1$.
We find then
\beq
\frac{\Omega_1}{2} \sigma_z^{(1)} +\frac{\Omega_2}{2} \sigma_z^{(2)}
&=& \frac{\Omega_1}{2} (P_+^{(1)} - P_-^{(1)}) (P_+^{(2)} +
P_-^{(2)})
+\frac{\Omega_2}{2} (P_+^{(1)} + P_-^{(1)}) (P_+^{(2)} - P_-^{(2)}) \nonumber \\
&=& \frac{\Omega_1 +\Omega_2}{2} P_{++} + \frac{\Omega_1 -
\Omega_2}{2} (P_{+-} - P_{-+}) - \frac{\Omega_1 + \Omega_2}{2}
P_{--}.
\eeq
It follows that the bare Green's function is 
\beq
G_0 (\omega) &=& \frac{P_{++}}{\omega - H_b - \frac{\Omega_1 + \Omega_2}{2} + i \eta} + \frac{P_{+-} }{\omega - H_b - \frac{\Omega_1 - \Omega_2}{2} + i \eta} \nonumber \\
&+&  \frac{P_{-+}}{\omega - H_b + \frac{\Omega_1 - \Omega_2}{2}+ i
\eta} +   \frac{P_{--}}{\omega - H_b +  \frac{\Omega_1 +
\Omega_2}{2} + i \eta} = P_{ab} G_{ab}.
\eeq

\subsection{Calculation of the self-energy and scattering matrices}
There is the only allowed diagram for the self-energy
\beq
& & \Sigma^{(2)} (\omega ) = a_{\nu} S_{+ \nu }  P_{ab} G_{ab}
(\omega) a_{\nu}^{\dagger}
   \begin{picture}(-20,11)
     \put(-80,8){\line(0,1){5}}
     \put(-80,13){\line(1,0){72}}
     \put(-8,8){\line(0,1){5}}
   \end{picture}
   \begin{picture}(20,11)
   \end{picture}
S_{- \nu} = S_{+ \nu }  P_{ab} G_{ab} (\omega - \nu) S_{- \nu} \nonumber \\
&=& \left[ g_1 \sigma_+^{(1)} e^{i \nu r_1} + g_2 \sigma_+^{(2)}
e^{i \nu r_2} \right] P_{ab} G_{ab}
\left[ g_1 \sigma_-^{(1)} e^{-i \nu r_1} + g_2 \sigma_-^{(2)} e^{-i \nu r_2} \right] \nonumber \\
&=&  \left[ g_1 e^{i \nu r_1} \sigma_+^{(1)}  P_{b}^{(2)} G_{-, b} + g_2 e^{i \nu r_2} P_{a}^{(1)} \sigma_+^{(2)} G_{a,-}\right] \left[ g_1 \sigma_-^{(1)} e^{-i \nu r_1} + g_2 \sigma_-^{(2)} e^{-i \nu r_2} \right] \nonumber \\
&=&  \int d \nu [ g_1^2 P_{+b} G_{-b} + g_1 g_2 e^{i \nu (r_2 -
r_1)} \sigma^{(1)}_- \sigma^{(2)}_+ G_{--} + g_1 g_2 e^{i \nu (r_1 -
r_2)} \sigma^{(1)}_+ \sigma^{(2)}_- G_{--} + g_2^2 P_{a,+} G_{a,-}]
\nonumber \\
&=& - i  \pi \left[ (g_1^2 + g_2^2) P_{++} + g_1^2 P_{+-} + g_2^2
P_{-+} + g_1 g_2 f (r_2 , r_1) \sigma^{(1)}_- \sigma^{(2)}_+ + g_1
g_2 f (r_1 , r_2) \sigma^{(1)}_+ \sigma^{(2)}_- \right],
\eeq
where $f (r_1 , r_2; \omega_b -H_b) =  f (r_1 , r_2) = 2 \exp[i (\omega_b -
H_b) (r_1 - r_2)] \Theta (r_1 - r_2)$, 
and $\omega = \omega_b - \Omega$, $\omega_b$ being the energy of
incoming (outgoing) photons. The standard symbol for the Wick contraction is also used. The dressed Green's function is therefore
\be
G (\omega) = G (\omega_b - H_b) = \frac{P_{++}}{\omega_b - H_b -
\Omega_1 - \Omega_2  + i \pi (g_1^2 + g_2^2)} +
\frac{P_{--}}{\omega_b - H_b  + i \eta} +\mathcal{M},
\label{Ggen}
\ee
where $\mathcal{M}$ is the matrix to be specified below.

Now we consider the two cases separately: the case of concentrated system (when positions of two emitters $r_{1}$ and $r_{2}$ coincide) and the case of distributed system when the coordinates of emitters are different. In the former case the analysis based on Eq.~\eq{eq:main-formula} is sufficient, while the latter case requires its extension.

\subsubsection{Concentrated case}
For coinciding positions of emitters $r_1 = r_2$ and we have $f_{12}=f_{21} =1$ and
\beq
& & \mathcal{M}=  \left\{ \left[ (G_{+-}^0)^{-1} + i \pi g_1^2 \right] P_{+-}
+\left[ (G_{-+}^0)^{-1} + i \pi g_2^2 \right] P_{-+} + i \pi g_1 g_2
\frac{\sigma^{(1)}_- \sigma^{(2)}_+
+ \sigma^{(1)}_+ \sigma^{(2)}_-}{2} \right\}^{-1} \nonumber \\
&=& \frac{1}{\det \mathcal{M}} \left\{ \left[ (G_{-+}^0)^{-1} + i \pi g_2^2
\right] P_{+-} +\left[ (G_{+-}^0)^{-1} + i \pi g_1^2 \right] P_{-+}
 - i \pi g_1 g_2 \frac{\sigma^{(1)}_- \sigma^{(2)}_+
+ \sigma^{(1)}_+ \sigma^{(2)}_-}{2} \right\},
\eeq
where
\beq
\det \mathcal{M} &=&  (G_{+-}^0)^{-1} (G_{-+}^0)^{-1} + i \pi \left[ g_2^2 (G_{+-}^0)^{-1} + g_1^2 (G_{-+}^0)^{-1}\right] \nonumber \\
&=& \left(\omega_b - H_b - \frac{\Omega_1 + \Omega_2}{2} - \Lambda_+ \right) \left(\omega_b - H_b - \frac{\Omega_1 + \Omega_2}{2} - \Lambda_- \right), \\
\Lambda_{\pm} &=& - \frac{i \pi (g_1^2 + g_2^2)}{2} \pm \frac12
\sqrt{- \pi^2 (g_1^2 + g_2^2)^2 + (\Omega_1 - \Omega_2)^2- 2 i \pi
(\Omega_1 - \Omega_2) (g_1^2 - g_2^2)}.
\eeq
In the limiting case $\Omega_1 = \Omega_2 = \Omega$
\beq
\mathcal{M} &=& \frac{P_{+-} + P_{-+}}{2} \left[ \frac{1}{\omega_b - H_b -
\Omega + i \pi (g_1^2 + g_2^2)} + \frac{1}{\omega_b - H_b - \Omega +
i \eta} \right]
\nonumber \\
&+& \frac{P_{+-} - P_{-+}}{2} \cos \theta \left[ \frac{1}{\omega_b -
H_b - \Omega + i \pi (g_1^2 + g_2^2)} - \frac{1}{\omega_b - H_b -
\Omega + i \eta} \right]
\nonumber \\
&+& \frac{\sigma^{(1)}_- \sigma^{(2)}_+ + \sigma^{(1)}_+
\sigma^{(2)}_-}{2} \sin \theta \left[ \frac{1}{\omega_b - H_b
-\Omega + i \pi (g_1^2 +g_2^2)} - \frac{1}{\omega_b - H_b -\Omega +
i \eta} \right] ,
\eeq
where $\tan \frac{\theta}{2} = \frac{g_2}{g_1}$. Then
\beq
T^{(1)} =  \left[ g_1 \sigma_-^{(1)} P^{(2)}_-
+ g_2 P_-^{(1)} \sigma_-^{(2)}  \right] G (\nu_1 ) \left[ g_1 \sigma_+^{(1)} P^{(2)}_- + g_2 P_-^{(1)} \sigma_+^{(2)}  \right] a_{\nu_1}^{\dagger} a_{\nu_2} 
= a_{\nu_1}^{\dagger} a_{\nu_2}   P_{--} \frac{g_1^2 +
g_2^2}{\nu_1 - \Omega + i \pi (g_1^2 + g_2^2)},
\eeq
and the corresponding scattering matrix is
\be
S_{pk} = \delta_{pk} \left( 1 - 2 \frac{i  \pi (g_1^2+g_2^2)}{p -
\Omega + i \pi (g_1^2 + g_2^2) }\right) =  \delta_{pk}
\frac{p-\Omega - i \pi (g_1^2 + g_2^2)}{p - \Omega + i \pi
(g_1^2+g_2^2)} .
\label{s1}
\ee

\subsubsection{Distributed system}
For distributed system $r_1 \neq r_2$ we obtain
\beq
\mathcal{M} &=& \frac{P_{+-} }{\omega_b - H_b - \Omega_1 + i \pi g_1^2} +
\frac{P_{-+}}{\omega_b - H_b - \Omega_2 + i \pi g_2^2}
\nonumber \\
&-& i \pi g_1 g_2 \frac{f (r_2 , r_1) \sigma^{(1)}_- \sigma^{(2)}_+
+ f (r_1 , r_2) \sigma^{(1)}_+ \sigma^{(2)}_-}{(\omega_b - H_b - \Omega_1 + i \pi g_1^2)(\omega_b - H_b - \Omega_2 + i \pi g_2^2)} \nonumber \\
&=& M_1 P_{+-} + M_2 P_{-+} + M_{12} \sigma^{(1)}_+ \sigma^{(2)}_- +
M_{21} \sigma^{(1)}_- \sigma^{(2)}_+ .
\label{Mpart}
\eeq
The one-photon $T$-matrix amounts to
\beq
T^{(1)} &=& P_{--}  a_{\nu_1}^{\dagger} S_{- , \nu_1} G a_{\nu_2} S_{+, \nu_2} P_{--} \nonumber \\
&=& \left[ g_1 \sigma_-^{(1)} P^{(2)}_- e^{- i \nu_1 r_1} + g_2
P_-^{(1)} \sigma_-^{(2)} e^{- i \nu_1 r_2}\right] G (\nu_1 ) \left[
g_1 \sigma_+^{(1)} P^{(2)}_- e^{i \nu_2 r_1} + g_2 P_-^{(1)}
\sigma_+^{(2)}  e^{i \nu_2 r_2}\right] a_{\nu_1}^{\dagger} a_{\nu_2}
\nonumber \\
&=&  a_{\nu_1}^{\dagger} a_{\nu_2}   P_{--} \frac{g_1^2 (\nu_1 -
\Omega_2) + g_2^2 (\nu_1 - \Omega_1 )}{(\nu_1 - \Omega_1 + i \pi
g_1^2) (\nu_1 - \Omega_2 + i \pi g_2^2)} ,
\eeq
and the corresponding scattering matrix reads
\be
S_{pk} = \delta_{pk} \left( 1 - 2  \frac{i \pi g_1^2 (p - \Omega_2 )
+ i \pi g_2^2 (p - \Omega_1 )}{(p - \Omega_1 + i \pi g_1^2) (p -
\Omega_2 + i \pi g_2^2)} \right) = \delta_{pk} \prod_{i=1}^2
\frac{(p - \Omega_i - i \pi g_i^2)}{(p - \Omega_i + i \pi g_i^2)} .
\label{s1s2onephoton}
\ee
We note that \eq{s1s2onephoton} can be represented as a convolution
\be
S_{pk} = \int d k' S_{2; pk'} S_{1; k'k}
\label{convol_onephot}
\ee
of the scattering matrices $S_1$ and $S_2$ on the first and the second emitters, respectively. This property is a consequence of the absent backscattering
for chiral photons, and it will be later on generalized to the arbitrary $N$-photon sector.

We also make the two following observations: 1) In the absence of backscattering there is no interference between counter-propagating waves, and therefore the outgoing state contains no information about positions of emitters, that is there is no dependence on $r_1 - r_2$ in \eq{s1s2onephoton}. 2) The result \eq{s1s2onephoton} would smoothly cross over to \eq{s1} on a scale of the phononic wavelength. The latter quantity is of the order of an inverse bandwidth, which is effectively set to zero in our theory. Therefore, \eq{s1} and \eq{s1s2onephoton}
are not analytically connected with each other in the limit of the vanishing distance $r_1 - r_2$ between the emitters.

\subsection{General approach: One-photon scattering}
Here we present a general approach aimed at calculation of the scattering matrices for the distributed system.
Therefore in the following we consider only the model with two emitters located
in different positions and having different $\Omega_i$ and $g_i$,
\beq
H  &=& \int d \nu \, \nu \, a^{\dagger} (\nu ) a (\nu ) +
\frac{\Omega_1}{2} \sigma_z^{(1)} + \frac{\Omega_2}{2}
\sigma_z^{(2)}  + V_1 + V_2 ,
\eeq
where $V_i = v_i + v_i^{\dagger}$,  where $i =1,2$. Here
$v_i = g_i \sigma^{(i)}_+ \int d \nu e^{i \nu r_i} a (\nu) \equiv
g_i \sigma^{(i)}_+ A_i $.
Let us label the atoms in such a way that $r_1 > r_2$. Our aim is to calculate the $T$-matrix
\be
T (\omega) =  V + V \frac{1}{\omega - H + i \eta} V ,
\label{Tmatrix}
\ee
where $V=V_1 +V_2$, in the ground state of the atomic system $|
\downarrow \downarrow \rangle$. In the following we omit $i \eta$
assuming $\omega \to \omega + i \eta$.

The first term in \eq{Tmatrix} can be neglected as it is
off-diagonal in spin states. In calculation of the second term
$(V_1 + V_2) \frac{1}{\omega - H_0 - V_1 + V_2} (V_1 +V_2)$
we can retain only $(v_1^{\dagger} + v_2^{\dagger})$ in the left
factor and $(v_1 + v_2)$ in the right factor, respectively, since
$\sigma_-^{i} |\downarrow \downarrow \rangle =0 $ and $\langle
\downarrow \downarrow | \sigma^{(i)}_+ =0$. Thus we have a sum of
four terms
\beq
T &=& v_1^{\dagger} \frac{1}{\omega - H_0 - V_1 - V_2} v_1 \label{tt11} \\
&+& v_2^{\dagger} \frac{1}{\omega - H_0 - V_1 - V_2} v_2 \label{tt22} \\
&+& v_1^{\dagger} \frac{1}{\omega - H_0 - V_1 - V_2} v_2 \label{tt12} \\
&+& v_2^{\dagger} \frac{1}{\omega - H_0 - V_1 - V_2} v_1 . \label{tt21}
\eeq
As a consequence of the RWA we obtain the operators $A_i^{\dagger}
\sim v_i^{\dagger}$ to the left from the resolvent, and the
operators $A_i \sim v_i$ to the right from the resolvent in each
term of this sum.

Let us now consider term by term expanding them first in $V_1$, and
then in $V_2$. In the following we will use an important observation
that $v_1$ can be only paired with the adjacent $v_1^{\dagger}$,
while any contraction of $v_2$ and $v_1^{\dagger}$ always yields
zero.  More generally, the contraction
\be
v_i v_j^{\dagger} \begin{picture}(-10,11)
     \put(-16,8){\line(0,1){5}}
     \put(-16,13){\line(1,0){9}}
     \put(-7,8){\line(0,1){5}}
   \end{picture}
   \begin{picture}(10,11)
   \end{picture}
 =0 , \quad i > j .
\label{contr_rule}
\ee
vanishes as it typically implies an integral of the kind
\be
\int d \nu \frac{e^{i \nu (r_i - r_j)}}{(\ldots - \nu + i \eta)
\ldots (\ldots - \nu + i \eta) } =0.
\ee
This integral is zero as it has all poles in the upper half-plane,
and the exponential function $e^{i \nu (r_i - r_j)}$ decays
sufficiently fast in the lower half-plane for $r_i < r_j$. Therefore
we can close the integration contour in the lower half-plane and get
zero.

Let us consider different components \eq{tt11}-\eq{tt21} of the $T$-matrix in more detail. In the contribution $T_{11}$ given by \eq{tt11}
\beq
T_{11} = v_1^{\dagger} \left( \frac{1}{\omega - H_0 - V_2} +
\frac{1}{\omega - H_0 - V_2} v_1 \frac{1}{\omega - H_0 - V_2}
v_1^{\dagger} \frac{1}{\omega - H_0 - V_2} + \ldots\right) v_1
\label{t11}
\eeq
we can omit the terms in parentheses which are odd in $V_1$, as they
are off-diagonal in the spin states of the $i=1$ atom. Moreover,
the operators  $v_1$ and $v_1^{\dagger}$ must alternate.

If we are exclusively interested in the single-photon scattering, then
we should only consider terms with the single creation (annihilation)
operator in the left (right) side.  For this reason we can neglect
$V_2$ everywhere in \eq{t11}: an expansion in $V_2$ cannot contain
$v_2$ in the leftmost position because of the spin state of the
second atom, and it cannot contain $v_2^{\dagger}$ because of the
photon state. Note that this argument is implicitly based on the
RWA.

Resumming the remaining series we obtain in the one-photon sector
\be
T_{11}^{(1)} = v_1^{\dagger} \frac{1}{\omega - H_0 - \Sigma_1} v_1 =
g_1^2 A_1^{\dagger} \frac{1}{\omega - H_0 - \Omega_1 - \Sigma_1}
A_1,
\ee
where $
\Sigma_1 = - i \pi g_1^2 \equiv  - i \Gamma_1
$.

Applying similar arguments to $T_{22}$ \eq{tt22}, we find
\be
T_{22}^{(1)} = v_2^{\dagger} \frac{1}{\omega - H_0 - \Sigma_2} v_2 =
g_2^2 A_2^{\dagger} \frac{1}{\omega - H_0 - \Omega_2 - \Sigma_2}
A_2,
\ee
where $
\Sigma_2 = - i \pi g_2^2 \equiv  - i \Gamma_2$.

Analogously we find for $T_{12}$ \eq{tt12} that
\beq
T_{12} = v_1^{\dagger} \left( \frac{1}{\omega - H_0 - V_2} v_1
\frac{1}{\omega - H_0 - V_2}   + \ldots\right) v_2 .
\label{t12}
\eeq
The terms in parentheses which are even in $V_1$ are omitted; an
expansion starts from $v_1$. Once a term containing $v_1^{\dagger}$
occurs (e.g. $\sim v_1 v_1^{\dagger} v_1$), a pairing of adjacent
$v_1$ and $v_1^{\dagger}$ should be performed, as there is no any
other possibility for $v_1^{\dagger}$ to be paired in the one-photon
sector (the latter implies that $v_1^{\dagger}$ can not be either
paired to an external vertex). This leads to an expression
\beq
T_{12}^{(1)} = v_1^{\dagger} \frac{1}{\omega - H_0 - V_2 - \Sigma_1}
v_1 \frac{1}{\omega - H_0 - V_2}  v_2 .
\label{t12-1}
\eeq
In the first propagator one can neglect $V_2$ as the corresponding
expansion can start neither form $v_2$ nor from $v_2^{\dagger}$.
After expanding the second propagator in $V_2$ one keeps only the
terms odd in $V_2$, the expansion starting from $v_2^{\dagger}$.
Resumming the series containing the power of contraction between
$v_2$ and $v_2^{\dagger}$, one transforms \eq{t12-1} into
\be
T_{12}^{(1)} = v_1^{\dagger} \frac{1}{\omega - H_0 - \Sigma_1} v_1
\frac{1}{\omega - H_0 } v_2^{\dagger} \frac{1}{\omega - H_0 -
\Sigma_2}  v_2 .
\label{t12-2}
\ee
It now only remains to pair $v_1$ and $v_2^{\dagger}$ which
results in
\be
T_{12}^{(1)} = - 2 \pi i g_1^2 g_2^2 A_1^{\dagger} \frac{1}{\omega -
H_0 -\Omega_1 -\Sigma_1}  e^{i (\omega - H_0) (r_1 - r_2) }
\frac{1}{\omega - H_0 - \Omega_2 -\Sigma_2}  A_2,
\label{t12-3}
\ee
where we have also used that
\be
\sigma_- \frac{1}{\omega - \frac{\Omega}{2} \sigma_z} =
\frac{1}{\omega - \frac{\Omega}{2} \sigma_z - \Omega} \sigma_- .
\ee

Finally, we consider $T_{21}$ \eq{tt21}
\beq
T_{21} = v_2^{\dagger} \left( \frac{1}{\omega - H_0 - V_2}
v_1^{\dagger} \frac{1}{\omega - H_0 - V_2}  + \ldots\right) v_1 .
\label{t21}
\eeq
The terms in parentheses which are even in $V_1$ are omitted; an
expansion starts from $v_1^{\dagger}$. However, this operator cannot
be paired to any operator standing to the left from it as follows
from \eq{contr_rule}. Therefore, $T_{21}$ identically vanishes in
the one-photon sector, $
T_{21}^{(1)} =0$.

Let us now collect all the terms and calculate the scattering matrix
in the one-photon sector
\beq
S_{pk} &=& \delta_{pk} - 2  i \delta_{pk}  \left[ \frac{\Gamma_1}{p - \Omega_1 + i \Gamma_1} + \frac{\Gamma_2}{p - \Omega_2 + i \Gamma_2}  - 2  i \Gamma_1 \Gamma_2 e^{- i p r_1} \frac{1}{p - \Omega_1 + i \Gamma_1 } e^{i p (r_1 - r_2)} \frac{1}{p -\Omega_2 + i \Gamma_2} e^{i p r_2} \right] \nonumber \\
&=& \delta_{pk} \left[ 1 - 2 \frac{i \Gamma_1 (p - \Omega_2) + i
\Gamma_2 (p - \Omega_1) }{(p - \Omega_1 + i \Gamma_1 ) (p -\Omega_2
+ i \Gamma_2 )}\right]
= \delta_{pk} \frac{(p - \Omega_1 - i \Gamma_1 ) (p - \Omega_2 - i
\Gamma_2 ) }{(p - \Omega_1 + i \Gamma_1 ) (p -\Omega_2 + i \Gamma_2
)} .
\eeq
which coincides with the result \eq{s1s2onephoton} from the previous subsection.

\subsection{Two-photon sector}
In order to find an exact expression for a scattering matrix in the
two-photon sector it is necessary to classify all possible
arrangements of external vertices with their eventual renormalization
which are allowed by the algebra of spin operators. In between
vertices one can insert the dressed Green's functions given by
Eqs.~\eq{Ggen} and \eq{Mpart}.

\begin{figure}[h]
\begin{center}
\includegraphics[width=12cm]{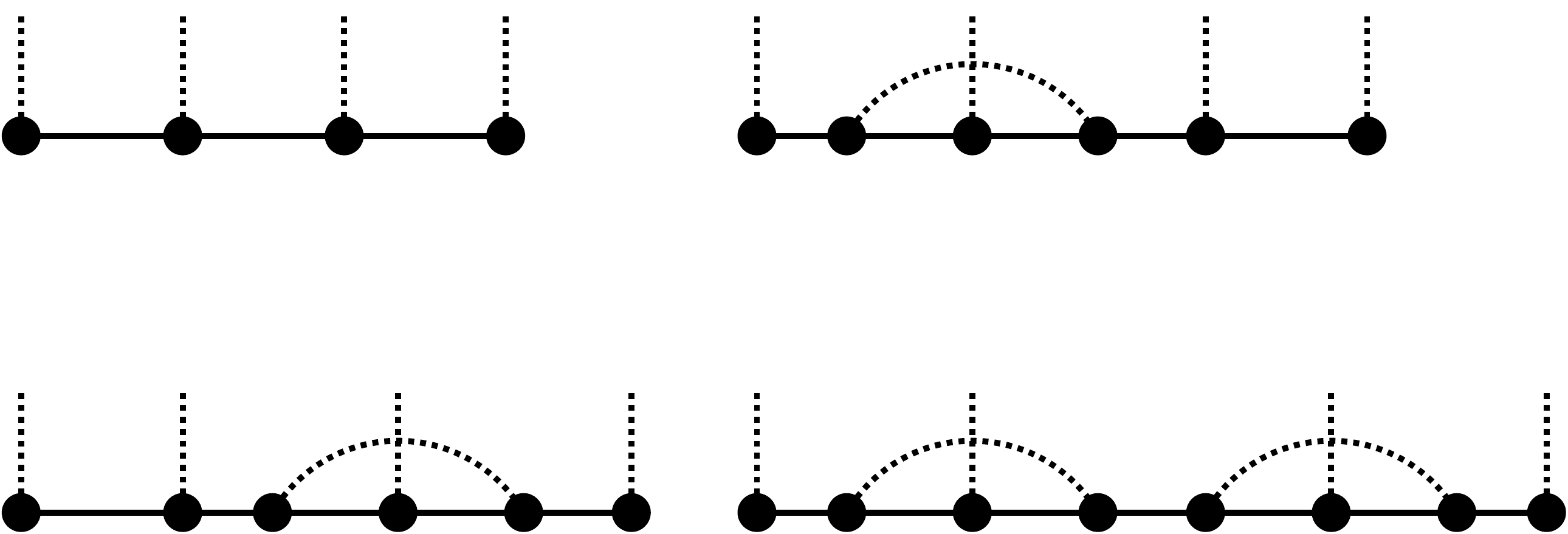}
\caption{Diagrams without renormalization of vertices and with renormalization of a single vertex. Solid lines correspond to the dressed Green's functions, circles denote spin operators, dotted lines are (paired and unpaired) photonic lines.}
\label{single_vert}
\end{center}
\end{figure}
\begin{figure}[h]
\begin{center}
\includegraphics[width=7cm]{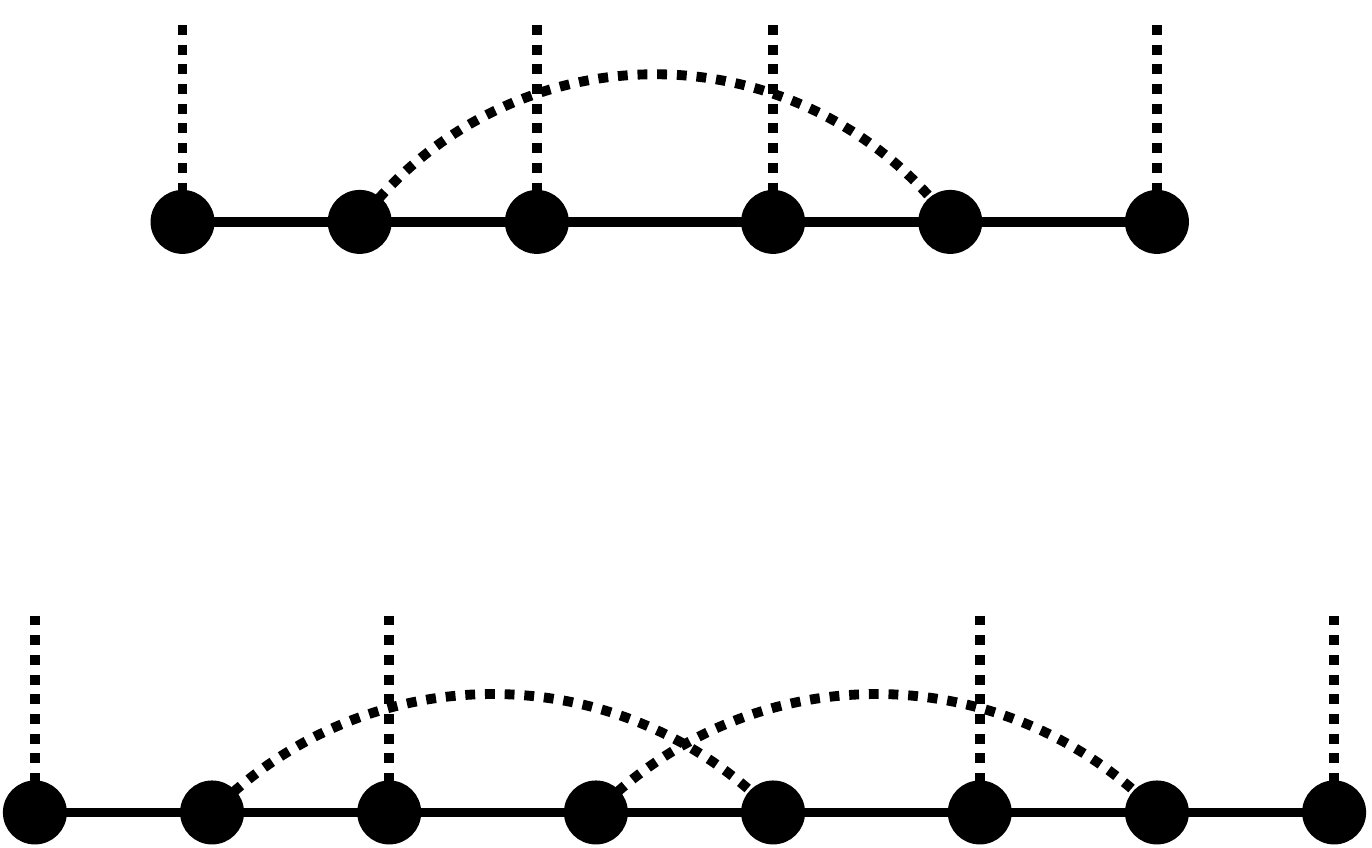}
\caption{Diagrams with renormalization of double vertices. Notations are the same as before.}
\label{double_vert}
\end{center}
\end{figure}

The diagrams which do not vanish in the two-photon sector are shown
in Figs.~\ref{single_vert} and \ref{double_vert}.
The first and the forth vertices in these diagrams are not renormalized, 
which is a
consequence of the RWA. This allows us to evaluate the part which is
the same for all diagrams
\beq
& & a^{\dagger}_{\nu_1} (g_1 \sigma_-^{(1)} P_-^{(2)} e^{- i \nu_1
r_1} + g_2 P_-^{(1)} \sigma_-^{(2)}  e^{- i \nu_1 r_2}) G \ldots G
a_{\nu_4} (g_1 \sigma_+^{(1)} P_-^{(2)} e^{i \nu_4 r_1} + g_2
P_-^{(1)} \sigma_+^{(2)}  e^{ i \nu_4 r_2})
\nonumber \\
&=& a^{\dagger}_{\nu_1} (g_1 M_1 \sigma_-^{(1)} P_-^{(2)} e^{- i \nu_1 r_1} + g_2 M_2 P_-^{(1)} \sigma_-^{(2)}  e^{- i \nu_1 r_2} + g_1 M_{12} P_-^{(1)} \sigma_-^{(2)} e^{-i \nu_1 r_1}) \times \nonumber \\
& & \times \ldots \times a_{\nu_4} (M_1 g_1 \sigma_+^{(1)} P_-^{(2)} e^{i \nu_4 r_1} + M_2 g_2 P_-^{(1)} \sigma_+^{(2)}  e^{ i \nu_4 r_2} +  M_{12} g_2 \sigma_+^{(1)} P_-^{(2)} e^{i \nu_4 r_2}) \nonumber \\
&=& a^{\dagger}_{\nu_1} (g_1 M_1 \sigma_-^{(1)} P_-^{(2)} e^{- i \nu_1 r_1} + g_2 M_2 S_1 P_-^{(1)} \sigma_-^{(2)}  e^{- i \nu_1 r_2} ) \nonumber \\
& & \times \ldots \times a_{\nu_4} (M_1 g_1 S_2 \sigma_+^{(1)}
P_-^{(2)} e^{i \nu_4 r_1} + M_2 g_2 P_-^{(1)} \sigma_+^{(2)}  e^{ i
\nu_4 r_2} ),
\eeq
where $S_{1,2} = 1- 2 \pi i g_{1,2}^2 M_{1,2}$, and $M_{1,2}$ are defined in \eq{Mpart}.

Let us first evaluate the diagrams shown in
Fig.~\ref{single_vert}. The diagram without vertex renormalization is given by  the expression
\beq
& &
P_{--} a^{\dagger}_{\nu_1} S_{- , \nu_1} G a_{\nu_2} S_{+ , \nu_2}  G a^{\dagger}_{\nu_3} S_{- , \nu_3}  G  a_{\nu_4} S_{+ , \nu_4} P_{--} \nonumber \\
&+& P_{--} a^{\dagger}_{\nu_1} S_{- , \nu_1} G a^{\dagger}_{\nu_2}
S_{- , \nu_2}  G a_{\nu_3} S_{+ , \nu_3}  G  a_{\nu_4} S_{+ , \nu_4}
P_{--} .
\label{t2pert}
\eeq
The first term in \eq{t2pert} reads
\beq
& & P_{--} a^{\dagger}_{\nu_1} \left( g_1^2 M_1 e^{- i (\nu_1 - \nu_2 ) r_1} + g_2^2 M_2 S_1 e^{- i (\nu_1 - \nu_2 ) r_2} \right) a_{\nu_2} \nonumber \\
& \times & \frac{1}{\omega_b - H_b + i \eta} a^{\dagger}_{\nu_3} \left( g_1^2 M_1 S_2 e^{- i (\nu_3 - \nu_4 ) r_1} + g_2^2 M_2 e^{- i (\nu_3 - \nu_4 ) r_2} \right) a_{\nu_4} \nonumber \\
&=& P_{--} \left( g_1^2 M_1 (\nu_1) e^{- i (\nu_1 - \nu_3 ) r_1} + g_2^2 M_2 (\nu_1) S_1 (\nu_1) e^{- i (\nu_1 - \nu_3 ) r_2} \right) \frac{1}{\nu_1 - \nu_3 + i \eta} \nonumber \\
& \times & \left( g_1^2 M_1 (\nu_4) S_2 (\nu_4) e^{ i (\nu_1 - \nu_3
) r_1} + g_2^2 M_2 (\nu_4) e^{i (\nu_1 - \nu_3 ) r_2}\right)
a^{\dagger}_{\nu_1}a^{\dagger}_{\nu_2} a_{\nu_3} a_{\nu_4} \nonumber \\
&=& \left( g_1^4 M_1 (\nu_1) M_1 (\nu_4 ) S_2 (\nu_4) + g_2^4  S_1 (\nu_1) M_2 (\nu_1) M_2 (\nu_4 ) \right. \nonumber \\
& & \left. + g_1^2 g_2^2 M_1 (\nu_1) M_2 (\nu_4) e^{-i (\nu_1 - \nu_3) (r_1 - r_2)}+ g_1^2 g_2^2 M_1 (\nu_4) S_2 (\nu_4 ) M_2 (\nu_1) S_1 (\nu_1) e^{i (\nu_1 - \nu_3) (r_1 - r_2)} \right) \nonumber \\
& \times & \frac{P_{--}}{\nu_1 - \nu_3 + i \eta}
a^{\dagger}_{\nu_1}a^{\dagger}_{\nu_2} a_{\nu_3} a_{\nu_4} .
\eeq
The second term in \eq{t2pert} reads
\beq
& & P_{--} a^{\dagger}_{\nu_1} g_1 g_2 (M_1 e^{- i \nu_1 r_1 - i \nu_2 r_2} + M_2 S_1 e^{- i \nu_1 r_2 - i \nu_2 r_1}) a^{\dagger}_{\nu_2} \nonumber \\
& \times & \frac{1}{\omega_b - H_b - \alpha_1 - \alpha_2 } a_{\nu_3} g_1 g_2 (M_1 S_2 e^{i \nu_3 r_2 + i \nu_4 r_1} + M_2 e^{i \nu_3 r_1 + i \nu_4 r_2}) a_{\nu_4} \\
&=& P_{--} g_1^2 g_2^2 \left( M_1 (\nu_1 ) e^{- i \nu_1 r_1 - i \nu_2 r_2} + M_2 (\nu_1) S_1 (\nu_1) e^{- i \nu_1 r_2 - i \nu_2 r_1} \right) \nonumber \\
& \times & \frac{1}{E - \alpha_1 - \alpha_2} (M_1 (\nu_4 ) S_2 (\nu_4 ) e^{i \nu_3 r_2 + i \nu_4 r_1} + M_2 (\nu_4) e^{i \nu_3 r_1 + i \nu_4 r_2})  a^{\dagger}_{\nu_1} a^{\dagger}_{\nu_2} a_{\nu_3} a_{\nu_4} \nonumber \\
&=& \left( M_1 (\nu_1 ) M_1 (\nu_4) S_2 (\nu_4) e^{- i (\nu_1 - \nu_4 ) (r_1 - r_2)} + M_2 (\nu_1 ) S_1 (\nu_1 ) M_2 (\nu_4) e^{i (\nu_1 - \nu_4) (r_1 - r_2)}  \right. \nonumber \\
& & \left. + M_2 (\nu_1 ) S_1 (\nu_1 ) M_1 (\nu_4) S_2 (\nu_4) e^{i (\nu_1 - \nu_3) (r_1-r_2)} + M_1 (\nu_1 ) M_2 (\nu_4 ) e^{- i (\nu_1 - \nu_3) (r_1- r_2)}\right) \nonumber \\
& \times & \frac{P_{--} g_1^2 g_2^2}{E - \alpha_1 - \alpha_2 }
a^{\dagger}_{\nu_1} a^{\dagger}_{\nu_2} a_{\nu_3} a_{\nu_4} ,
\label{non_ren}
\eeq
where $\alpha_{1,2} = \Omega_{1,2} - i \pi g^2_{1,2}$ and $E = \nu_1
+ \nu_2 = \nu_3 + \nu_4$.

In order to evaluate the diagram with renormalization of the second
vertex we find the following  vertex correction to the second vertex
\beq
& & \left( g_1 \sigma_+^{(1)} a_{\nu} e^{i \nu r_1} \right) G \left( g_1 \sigma_-^{(1)} a_{\nu_2}^{\dagger} e^{- i \nu_2 r_1} \right) G \left( g_2 \sigma_-^{(2)} a_{\nu}^{\dagger} e^{-i \nu r_2} \right) \nonumber \\
&=& g_1^2 g_2 a_{\nu} G_{--}  a_{\nu_2}^{\dagger} G_{+-}
a_{\nu}^{\dagger}
\begin{picture}(-75,11)
     \put(-72,8){\line(0,1){5}}
     \put(-72,13){\line(1,0){65}}
     \put(-7,8){\line(0,1){5}}
   \end{picture}
   \begin{picture}(75,11)
   \end{picture}
P_+^{(1)} \sigma_-^{(2)} e^{i \nu (r_1 - r_2)} e^{- i \nu_2 r_1} \nonumber \\
&=& P_+^{(1)} \sigma_-^{(2)} e^{- i \nu_2 r_1} g_1^2 g_2 \int d \nu
\frac{e^{i \nu (r_1 - r_2)}}{(\omega_b - H_b -\nu + i \eta)
(\omega_b - H_b - \nu + \nu_2 - \alpha_1) } a_{\nu_2}^{\dagger}.
\eeq
Such vertex correction can only occur in the diagram which has
$v_1^{\dagger}$ in the first position (from the left). Therefore we
can effectively replace $\omega_b - H_b \to \nu_1$. Evaluating the
integral we obtain
\beq
& & P_+^{(1)} \sigma_-^{(2)} e^{- i \nu_2 r_1} g_1^2 g_2 (- 2 \pi i) e^{i \nu_1 (r_1 - r_2)} M_1 (\nu_2) \left[ 1 - e^{i (\nu_2 - \alpha_1) (r_1 - r_2)}\right]  a_{\nu_2}^{\dagger} \nonumber \\
&=& P_+^{(1)} \sigma_-^{(2)} e^{- i \nu_2 r_1} g_2 e^{i \nu_1 (r_1 -
r_2)} [ S_1 (\nu_2) -1] \left[ 1 - e^{i (\nu_2 - \alpha_1) (r_1 -
r_2)}\right]  a_{\nu_2}^{\dagger} .
\eeq
Now we can evaluate the whole diagram with this vertex correction
\beq
& & M_1 (\nu_1)  \left[ S_1 (\nu_2) -1 \right] \left[ 1 - e^{i (\nu_2 - \alpha_1) (r_1 - r_2)} \right] \nonumber \\
& \times & (M_1 (\nu_4 ) S_2 (\nu_4 ) e^{i (\nu_1 - \nu_3 ) (r_1 - r_2)} + M_2 (\nu_4) e^{i (\nu_1 - \nu_4) (r_1 - r_2 )})   \nonumber \\
& \times & \frac{P_{--} g_1^2 g_2^2 }{E - \alpha_1 - \alpha_2}
a^{\dagger}_{\nu_1} a^{\dagger}_{\nu_2} a_{\nu_3} a_{\nu_4}.
\label{ren_2}
\eeq

In order to evaluate the diagram with renormalization of the third
vertex we find the following vertex correction to the third vertex
\beq
& & \left( g_1 \sigma_+^{(1)} a_{\nu} e^{i \nu r_1} \right) G \left( g_2 \sigma_+^{(2)} a_{\nu_3} e^{ i \nu_3 r_2} \right) G \left( g_2 \sigma_-^{(2)} a_{\nu}^{\dagger} e^{-i \nu r_2} \right) \nonumber \\
&=& g_1 g_2^2 a_{\nu} G_{-+}  a_{\nu_3} G_{--} a_{\nu}^{\dagger}
\begin{picture}(-75,11)
     \put(-72,8){\line(0,1){5}}
     \put(-72,13){\line(1,0){65}}
     \put(-7,8){\line(0,1){5}}
   \end{picture}
   \begin{picture}(75,11)
   \end{picture}
\sigma_+^{(1)} P_+^{(2)} e^{i \nu (r_1 - r_2)} e^{i \nu_3 r_2} \nonumber \\
&=& \sigma_+^{(1)} P_+^{(2)} e^{ i \nu_3 r_2} g_1 g_2^2 a_{\nu_3}
\int d \nu \frac{e^{i \nu (r_1 - r_2)}}{(\omega_b - H_b -\nu + \nu_3
- \alpha_2 ) (\omega_b - H_b - \nu  + i \eta) } .
\eeq
Such vertex correction can only occur in the diagram which has $v_2$
in the forth position (from the left). Therefore we can effectively
replace $\omega_b - H_b \to \nu_4$. Evaluating the integral we
obtain
\beq
& & \sigma_+^{(1)} P_+^{(2)} e^{ i \nu_3 r_2} g_1 g_2^2 a_{\nu_3} (- 2 \pi i) e^{i \nu_4 (r_1 - r_2)} M_2 (\nu_3) \left[ 1 - e^{i (\nu_3 - \alpha_2) (r_1 - r_2)}\right] \nonumber \\
&=&    \sigma_+^{(1)} P_+^{(2)}  e^{ i \nu_3 r_2} g_1 e^{i \nu_4
(r_1 - r_2)} \left[ S_2 (\nu_3 ) -1 \right] \left[ 1 - e^{i (\nu_3 -
\alpha_2) (r_1 - r_2)}\right] a_{\nu_3} .
\eeq
Now we can evaluate the whole diagram with this vertex correction
\beq
& &  \left( M_1 (\nu_1 ) e^{- i (\nu_1 - \nu_4) (r_1 - r_2 ) } + M_2 (\nu_1) S_1 (\nu_1) e^{i (\nu_1 - \nu_3) (r_1 - r_2 )} \right) \nonumber \\
&\times & \left[ S_2 (\nu_3 ) -1 \right] \left[ 1 - e^{i (\nu_3 -
\alpha_2) (r_1 - r_2)}\right]  M_2 (\nu_4)  \frac{P_{--} g_1^2 g_2^2
}{E - \alpha_1 - \alpha_2} a_{\nu_1}^{\dagger } a_{\nu_2}^{\dagger}
a_{\nu_3} a_{\nu_4}.
\label{ren_3}
\eeq

The diagram with renormalization of both the second and the third
vertices is given by
\beq
& &     e^{ i (\nu_1 -\nu_3 ) (r_1 - r_2)}  \left[ 1 - e^{i (\nu_2 -
\alpha_1) (r_1 - r_2)}\right] \left[ 1 - e^{i (\nu_3 - \alpha_2)
(r_1 - r_2)}\right]
\nonumber \\
&\times &  M_1 (\nu_1) [ S_1 (\nu_2) -1] \left[ S_2 (\nu_3 ) -1
\right]   M_2 (\nu_4)  \frac{P_{--} g_1^2 g_2^2 }{E - \alpha_1 -
\alpha_2 } a_{\nu_1}^{\dagger } a_{\nu_2}^{\dagger} a_{\nu_3}
a_{\nu_4}.
\label{ren_23}
\eeq

Summarizing the results of Eqs.~\eq{non_ren}, \eq{ren_2},
\eq{ren_3}, and \eq{ren_23} we obtain the following intermediate
expression for $T^{(2)}$
\beq
&-&  i \pi \delta_{\nu_1 \nu_3} \left[ g_1^4 M_1 (\nu_1) M_1 (\nu_4 ) S_2 (\nu_4) + g_2^4  S_1 (\nu_1) M_2 (\nu_1) M_2 (\nu_4 ) \right]  \nonumber \\
&-&   i \pi \delta_{\nu_1 \nu_3} g_1^2 g_2^2 \left[ M_1 (\nu_1) M_2
(\nu_4) +
M_1 (\nu_4) S_2 (\nu_4) M_2 (\nu_1) S_1 (\nu_1 )  \right] \nonumber \\
&+& \frac{1}{\nu_1 - \nu_3} \left[ g_1^4 M_1 (\nu_1) M_1 (\nu_4 ) S_2 (\nu_4) + g_2^4  S_1 (\nu_1) M_2 (\nu_1) M_2 (\nu_4 ) \right]  \nonumber \\
&+& \frac{g_1^2 g_2^2}{\nu_1 - \nu_3} \left[ M_1 (\nu_1) M_2 (\nu_4) -  M_1 (\nu_3) S_2 (\nu_3 ) M_2 (\nu_2) S_1 (\nu_2) \right] e^{-i (\nu_1 - \nu_3) (r_1 - r_2)} \nonumber \\
&+& \frac{g_1^2 g_2^2}{E - \alpha_1 - \alpha_2} e^{-i (\nu_1 - \nu_3) (r_1 - r_2)} \nonumber \\
& \times & \left\{ M_1 (\nu_1)  + M_2 (\nu_2) S_1 (\nu_2) + M_1 (\nu_2) [ S_1 (\nu_1 ) -1 ] [1 - e^{i (\nu_1 - \alpha_1) (r_1 - r_2)}]  \right\} \nonumber \\
& \times & \left\{ M_2 (\nu_4)  + M_1 (\nu_3) S_2 (\nu_3) + M_2
(\nu_3) [ S_2 (\nu_4 ) -1 ] [1 - e^{i (\nu_4 - \alpha_2) (r_1 -
r_2)}]  \right\} ,
\label{intermed}
\eeq
which has to be convoluted with $a_{\nu_1}^{\dagger }
a_{\nu_2}^{\dagger} a_{\nu_3} a_{\nu_4}$. The projector $P_{--}$
onto the ground state of the atomic system is also omitted.

We see that the dependence on atomic positions is still present in
\eq{intermed}. Exchanging the dummy frequencies $\nu_1
\leftrightarrow \nu_2$ and $\nu_3 \leftrightarrow \nu_4$, when
necessary, and exploiting obvious identities
\beq
M_1 (\nu_2) [ S_1 (\nu_1 ) -1 ] &=& M_1 (\nu_1) [ S_1 (\nu_2 ) -1 ], \\
M_2 (\nu_3) [ S_2 (\nu_4 ) -1 ] &=& M_2 (\nu_4) [ S_2 (\nu_3 ) -1 ],
\eeq
we can cast the position-dependent part of \eq{intermed} to
\beq
& &  \frac{g_1^2 g_2^2}{\nu_1 - \nu_3} M_1 (\nu_1) M_2 (\nu_4) \left[1 - S_1 (\nu_2) S_2 (\nu_3)\right] e^{- i (\nu_1 - \nu_3) (r_1 - r_2)} \nonumber \\
&-& g_1^2 g_2^2 M_1 (\nu_1) [S_1 (\nu_2) - 1] M_1 (\nu_3) M_2 (\nu_4) S_2 (\nu_3) e^{i (\nu_3 - \alpha_1 ) (r_1 - r_2 )} \nonumber \\
&-& g_1^2 g_2^2 M_2 (\nu_4) [ S_2 (\nu_3 ) -1 ] M_1 (\nu_1) M_2 (\nu_2) S_1 (\nu_2 ) e^{i (\nu_2 - \alpha_2) (r_1 - r_2 )} \nonumber \\
&+& \frac{g_1^2 g_2^2}{E - \alpha_1 - \alpha_2} M_1 (\nu_1 ) M_2
(\nu_4 ) [S_1 (\nu_2) -1] [S_2 (\nu_3 ) -1] e^{i (E - \alpha_1 -
\alpha_2) (r_1 - r_2)} .
\label{pos_dep}
\eeq

A cancellation of \eq{pos_dep} is achieved by an account of
diagrams containing double vertices shown in Fig.~\ref{double_vert}.
The first possibility to generate a double vertex is provided by one
photon line covering the second and third vertices (upper panel of
Fig.~\ref{double_vert}). The corresponding double vertex reads
\beq
& & \left( g_1 \sigma_+^{(1)} a_{\nu} e^{i \nu r_1} \right) G \left(
g_i \sigma_-^{(i)} a_{\nu_2}^{\dagger} e^{-i \nu_2 r_i} \right) G
\left( g_i \sigma_+^{(i)} a_{\nu_3} e^{i \nu_3 r_i}\right) G \left(
g_2 \sigma_-^{(2)} a_{\nu}^{\dagger}
e^{-i \nu r_2}\right) \nonumber \\
&=& g_1^3 g_2 \sigma_+^{(1)} \sigma_-^{(2)} a_{\nu} G_{--}
a_{\nu_2}^{\dagger} G_{+-} a_{\nu_3} G_{--} a_{\nu}^{\dagger}
   \begin{picture}(-110,11)
     \put(-107,8){\line(0,1){5}}
     \put(-107,13){\line(1,0){100}}
     \put(-7,8){\line(0,1){5}}
   \end{picture}
   \begin{picture}(110,11)
   \end{picture}
e^{i \nu (r_1 - r_2)} e^{- i (\nu_2 - \nu_3) r_1}\nonumber \\
&+& g_1 g_2^3 \sigma_+^{(1)} \sigma_-^{(2)} a_{\nu} G_{--}
a_{\nu_2}^{\dagger} G_{-+} a_{\nu_3} G_{--} a_{\nu}^{\dagger}
  \begin{picture}(-110,11)
     \put(-107,8){\line(0,1){5}}
     \put(-107,13){\line(1,0){100}}
     \put(-7,8){\line(0,1){5}}
   \end{picture}
   \begin{picture}(110,11)
   \end{picture}
e^{i \nu (r_1 - r_2)} e^{- i (\nu_2 - \nu_3) r_2} \nonumber \\
&=& - g_1 g_2  \sigma_+^{(1)} \sigma_-^{(2)} a_{\nu_2}^{\dagger}
a_{\nu_3}
\left\{ g_1^2 e^{- i (\nu_2 - \nu_3) r_1} \int d \nu \frac{e^{i \nu (r_1 - r_2)}}{(\nu-\nu_1 - i \eta )(\nu- E + \alpha_1) (\nu-\nu_4 - i \eta )} \right. \nonumber \\
& & \left. + g_2^2 e^{- i (\nu_2 - \nu_3) r_2} \int d \nu \frac{e^{i \nu (r_1 - r_2)}}{(\nu -\nu_1 - i \eta )(\nu- E  + \alpha_2) (\nu -\nu_4 - i \eta )} \right\} \nonumber \\
&=&  2 \pi i g_1 g_2  \sigma_+^{(1)} \sigma_-^{(2)} a_{\nu_2}^{\dagger} a_{\nu_3} \\
& \times & \left\{ g_1^2 e^{- i (\nu_2 - \nu_3) r_1} \left[  \frac{ M_1 (\nu_2) e^{i \nu_1 (r_1 - r_2) }- M_1 (\nu_3 )  e^{i \nu_4 (r_1 - r_2) } }{\nu_1 - \nu_4} - M_1 (\nu_2) M_1 (\nu_3) e^{i (E - \alpha_1) (r_1 - r_2)}  \right] \right. \nonumber \\
& & \left. + g_2^2 e^{- i (\nu_2 - \nu_3) r_2} \left[  \frac{ M_2
(\nu_2) e^{i \nu_1 (r_1 - r_2) }- M_2 (\nu_3 )  e^{i \nu_4 (r_1 -
r_2) } }{\nu_1 - \nu_4} - M_2 (\nu_2) M_2 (\nu_3) e^{i (E -
\alpha_2) (r_1 - r_2)}  \right] \right\} \nonumber.
\eeq
The whole diagram amounts to
\beq
& &  g_1^2 g_2^2   a_{\nu_1}^{\dagger} a_{\nu_2}^{\dagger} a_{\nu_3}
a_{\nu_4}
\nonumber  \\
& \times & \left\{ [S_1 (\nu_2) -1 ] M_2 (\nu_4)\frac{ M_1 (\nu_1) e^{-i (\nu_1 - \nu_3 ) (r_1 - r_2) }}{\nu_1 - \nu_3} + [S_1 (\nu_1) -1 ] M_2 (\nu_3)\frac{M_1 (\nu_4 ) }{\nu_1 - \nu_3}    \right. \nonumber \\
& & + M_1 (\nu_1) M_2 (\nu_4) [ S_1 (\nu_2) -1 ] M_1 (\nu_3) e^{i (\nu_3 - \alpha_1) (r_1 - r_2)} \nonumber \\
& & +  M_1 (\nu_2) [S_2 (\nu_4)-1] \frac{ M_2 (\nu_1)}{\nu_1 - \nu_3} + M_1 (\nu_1) [ S_2 (\nu_3) - 1] \frac{M_2 (\nu_4 )  e^{-i (\nu_1 - \nu_3 ) (r_1 - r_2) } }{\nu_1 - \nu_3} \nonumber \\
& & \left.  + M_1 (\nu_1) M_2 (\nu_4) M_2 (\nu_2) [ S_2 (\nu_3) -1]
e^{i (\nu_2 - \alpha_2) (r_1 - r_2)}  \right\} .
\label{double_direct_diag}
\eeq

The second possibility to generate a double vertex is provided by
two photon lines covering the second and the third vertices,
respectively, and at the same time intersecting each other (lower
panel of of Fig.~\ref{double_vert}). The corresponding expression
reads
\beq
& & \left( g_1 \sigma_+^{(1)} a_{\nu} e^{i \nu r_1} \right) G \left(  g_1 \sigma_-^{(1)} a_{\nu_2}^{\dagger} e^{-i \nu_2 r_1} \right) G \left( g_1 \sigma_+^{(1)} a_{\mu} e^{i \mu r_1}\right) \nonumber \\
& \times & G \left( g_2 \sigma_-^{(2)} a_{\nu}^{\dagger} e^{-i \nu r_2}\right) G \left( g_2 \sigma_+^{(2)} a_{\nu_3} e^{i \nu_3 r_2}\right) G \left( g_2 \sigma_-^{(2)} a_{\mu}^{\dagger} e^{-i \mu r_2}\right) \nonumber \\
&=& g_1^3 g_2^3 \sigma_+^{(1)} \sigma_-^{(2)} a_{\nu} G_{--}
a_{\nu_2}^{\dagger} G_{+-} a_{\mu} G_{--}  a_{\nu}^{\dagger}
\begin{picture}(-110,11)
     \put(-104,8){\line(0,1){3}}
     \put(-104,11){\line(1,0){97}}
     \put(-7,8){\line(0,1){3}}
   \end{picture}
   \begin{picture}(110,11)
   \end{picture}
G_{-+} a_{\nu_3} G_{--} a_{\mu}^{\dagger}
\begin{picture}(-110,11)
     \put(-105,8){\line(0,1){5}}
     \put(-105,13){\line(1,0){98}}
     \put(-7,8){\line(0,1){5}}
   \end{picture}
   \begin{picture}(110,11)
   \end{picture}
e^{i \nu (r_1 - r_2)} e^{i \mu (r_1 - r_2)} e^{- i \nu_2 r_1 + i
\nu_3 r_2}
\nonumber \\
&=& - g_1^3 g_2^3 e^{- i \nu_2 r_1 + i \nu_3 r_2} \sigma_+^{(1)}
\sigma_-^{(2)}
a_{\nu_2}^{\dagger} a_{\nu_3} \nonumber \\
& \times &\int d \nu \int d \mu \frac{e^{i \nu (r_1 - r_2)} e^{i \mu
(r_1 - r_2)}}{(\nu - \nu_1 - i \eta ) (\nu - E + \alpha_1) (\nu +
\mu - E - i \eta) (\mu - E +\alpha_2) (\mu - \nu_4 - i \eta )}.
\label{double_entang}
\eeq

Let us first evaluate the integral over $\mu$
\beq
& & \int  d \mu \frac{e^{i \mu (r_1 - r_2)}}{ (\nu + \mu - E - i \eta) (\mu - E +\alpha_2) (\mu - \nu_4 - i \eta )} \nonumber \\
&=& 2 \pi i \left[ \frac{e^{i (E - \nu) (r_1 - r_2)}}{(- \nu + \nu_2) (\nu_3 - \nu )} + \frac{e^{i (E - \alpha_2) (r_1 - r_2)}}{(\nu - \alpha_2 ) (\nu_3 - \alpha_2)} + \frac{e^{i \nu_4 (r_1 - r_2)}}{(\nu - \nu_3) (- \nu_3 + \alpha_2 )} \right] \nonumber \\
&=& 2 \pi i e^{i (E-\nu ) (r_1 - r_2)} M_2 (\nu_3) \left[ \frac{1 -
e^{i (\nu - \nu_3) (r_1 - r_2)}}{\nu - \nu_3} - \frac{1- e^{i (\nu -
\alpha_2) (r_1 - r_2) }}{\nu - \alpha_2} \right].
\eeq
Now we have to perform integration over $\nu$. Note that there is no
pole at $\nu = \nu_3$ and $\nu =\alpha_2$. Collecting the
contributions from the poles at $\nu = \nu_1 + i \eta$ and $\nu =
E-\alpha_1$, we cast \eq{double_entang} to
\beq
& & (2 \pi i)^2 g_1^3 g_2^3 e^{i \nu_1 r_1 - i \nu_4 r_2} \sigma_+^{(1)} \sigma_-^{(2)} a_{\nu_2}^{\dagger} a_{\nu_3} M_2 (\nu_3) M_1 (\nu_2 ) \nonumber \\
& \times & \left\{ \frac{1 - e^{i (\nu_1 - \nu_3) (r_1 - r_2)}}{\nu_1 - \nu_3}  + \frac{1- e^{i (E - \alpha_1 - \alpha_2) (r_1 - r_2) }}{E - \alpha_1  - \alpha_2} \right. \nonumber \\
& & \left. - M_2 (\nu_1 ) [1- e^{i (\nu_1 - \alpha_2) (r_1 - r_2) }]
-  M_1 (\nu_4) [1 - e^{i (\nu_4 - \alpha_1) (r_1 - r_2)}] \right\} .
\eeq
The whole diagram with this vertex amounts to
\beq
& & (2 \pi i)^2 g_1^4 g_2^4 P_{--} a_{\nu_1}^{\dagger} a_{\nu_2}^{\dagger} a_{\nu_3} a_{\nu_4} M_1 (\nu_1) M_2 (\nu_3) M_1 (\nu_2 ) M_2 (\nu_4) \nonumber \\
& \times & \left\{ \frac{1 - e^{i (\nu_1 - \nu_3) (r_1 - r_2)}}{\nu_1 - \nu_3}  + \frac{1- e^{i (E - \alpha_1 - \alpha_2) (r_1 - r_2) }}{E - \alpha_1  - \alpha_2} \right. \nonumber \\
& & \left. - M_2 (\nu_1 ) [1- e^{i (\nu_1 - \alpha_2) (r_1 - r_2) }]  -  M_1 (\nu_4) [1 - e^{i (\nu_4 - \alpha_1) (r_1 - r_2)}] \right\}  \nonumber \\
&=& g_1^2 g_2^2 P_{--} a_{\nu_1}^{\dagger} a_{\nu_2}^{\dagger} a_{\nu_3} a_{\nu_4} \left\{ - M_1 (\nu_2) M_2 (\nu_4) [S_1 (\nu_1 ) -1] [S_2 (\nu_3) -1]\frac{1 - e^{-i (\nu_1 - \nu_3) (r_1 - r_2)}}{\nu_1 - \nu_3}\right. \nonumber \\
& & + M_1 (\nu_1) [S_2 (\nu_3)-1] [S_1 (\nu_2 ) -1] M_2 (\nu_4) \frac{1- e^{i (E - \alpha_1 - \alpha_2) (r_1 - r_2) }}{E - \alpha_1  - \alpha_2} \nonumber \\
& & -  [S_1 (\nu_2) -1] [S_2 (\nu_3)-1] M_1 (\nu_1 ) M_2 (\nu_4) M_2 (\nu_2 ) [1- e^{i (\nu_2 - \alpha_2) (r_1 - r_2) }] \nonumber \\
& & \left. - M_1 (\nu_1) M_2 (\nu_4) [S_1 (\nu_2 ) -1] [S_2 (\nu_3)
-1] M_1 (\nu_3) [1 - e^{i (\nu_3 - \alpha_1) (r_1 - r_2)}]\right\} .
\label{double_entang_diag}
\eeq

Collecting all contributions to $T^{(2)}$, namely \eq{pos_dep},
\eq{double_direct_diag}, \eq{double_entang_diag}, and the terms
$\sim g_1^4 , g_2^4$ from \eq{intermed}, we obtain
\beq
-2 \pi i T^{(2)} &=& - \frac{2 \pi i}{\nu_1 - \nu_3} \left[ g_1^4 M_1 (\nu_1) M_1 (\nu_4 ) S_2 (\nu_4) + g_2^4  S_1 (\nu_1) M_2 (\nu_1) M_2 (\nu_4 ) \right]  \nonumber \\
&-&  2 \pi i g_1^2 g_2^2 \left\{  [S_1 (\nu_1) -1 ] M_2 (\nu_3)\frac{M_1 (\nu_4 ) }{\nu_1 - \nu_3}  +  M_1 (\nu_2) [S_2 (\nu_4)-1] \frac{ M_2 (\nu_1)}{\nu_1 - \nu_3}  \right. \nonumber \\
& &   - M_1 (\nu_2) M_2 (\nu_4)  [S_1 (\nu_1 ) -1]
[S_2 (\nu_3)-1] \frac{1}{\nu_1 - \nu_3}  \nonumber \\
& & + M_1 (\nu_1) [S_2 (\nu_3)-1] [S_1 (\nu_2 ) -1] M_2 (\nu_4) \frac{1}{E - \alpha_1  - \alpha_2} \nonumber \\
& & -  [S_1 (\nu_2) -1] [S_2 (\nu_3) -1] M_1 (\nu_1 ) M_2 (\nu_4) M_2 (\nu_2 )  \nonumber \\
& & \left. - M_1 (\nu_1) M_2 (\nu_4) [S_1 (\nu_2 ) -1] [S_2 (\nu_3)
-1] M_1 (\nu_3) \right\}.
\eeq

One can observe the dependence on atomic coordinates has
disappeared. Transforming $T^{(2)}$ further, we find
\beq
-2 \pi i T^{(2)} &=&  \frac{E - 2 \alpha_2}{4 \pi i}  S_1 (\nu_1 )  S_1 (\nu_2)   M_2 (\nu_1) M_2 (\nu_2) [S_2 (\nu_4) -1]  [S_2 (\nu_3)-1]   \nonumber \\
&+&  \frac{E- 2 \alpha_1}{4 \pi i} M_1 (\nu_4) M_1 (\nu_3 ) [S_1 (\nu_2)-1]   S_2 (\nu_4) [S_1 (\nu_1 ) -1]  S_2 (\nu_3) \nonumber \\
&-& \frac{1}{2 \pi i}  [S_1 (\nu_1 ) -1] [S_2 (\nu_3)-1] [S_1 (\nu_2
) -1] [S_2 (\nu_4) -1] \frac{1}{E - \alpha_1  - \alpha_2} .
\label{T2fin}
\eeq
This expression should be multiplied by $a^{\dagger}_{\nu_1}
a^{\dagger}_{\nu_2} a_{\nu_3} a_{\nu_4} $. Then the irreducible part
of $S^{(2)}$ equals
\be
S^{(2)irred}_{p_1 p_2, k_1 k_2} = -2 \pi i \delta_{p_1 + p_2 , k_1
+k_2} \langle a_{p_2} a_{p_1} T^{(2)} a_{k_1}^{\dagger}
a_{k_2}^{\dagger} \rangle .
\label{s2irred}
\ee
Note that if one of the atoms is decoupled (say, $g_2 = 0$), then
$S_2 \equiv 1$, and one recovers from \eq{T2fin} the already known
result for the two-photon scattering matrix on a single atom.

It now remains to evaluate a reducible contribution to the
two-photon scattering.  To this end we use an expression for
$T^{(1)}$ and the $\delta$-part of \eq{pos_dep} and find that the
corresponding expression factorizes into products of the one-photon
scattering matrices
\beq
S^{(2)red}_{p_1 p_2, k_1 k_2} &=&  (\delta_{p_1 k_1} \delta_{p_2 k_2
}+ \delta_{p_1 k_2} \delta_{p_2 k_1 } )\left\{ 1- 2 \pi i  \left[
g_1^2 M_1 (p_2) + g_2^2 M_2 (p_2) - 2 \pi i g_1^2 g_2^2 M_1 (p_2 )
M_2 (p_2)\right] \right.
\nonumber \\
& & - 2 \pi i  \left[ g_1^2 M_1 (p_1) + g_2^2 M_2 (p_1)  -2 \pi i g_1^2 g_2^2 M_1 (p_1 ) M_2 (p_1)\right] \nonumber \\
& &  + \frac12 (2 \pi i)^2   \left[ g_1^4 M_1 (p_1) M_1 (p_2) S_2 (p_2) + g_2^4 S_1 (p_1) M_2 (p_1) M_2 (p_2)\right]   \nonumber \\
& &  + \frac12 (2 \pi i)^2 g_1^2 g_2^2 \left[ M_1 (p_1) M_2 (p_2) + M_1 (p_2) S_2 (p_2) M_2 (p_1) S_1 (p_1)\right]   \nonumber \\
& &  + \frac12 (2 \pi i)^2 \left[ g_1^4 M_1 (p_2) M_1 (p_1) S_2 (p_1) + g_2^4 S_1 (p_2) M_2 (p_2) M_2 (p_1)\right]   \nonumber \\
& & \left. + \frac12 (2 \pi i)^2 g_1^2 g_2^2 \left[ M_1 (p_2) M_2 (p_1) + M_1 (p_1) S_2 (p_1) M_2 (p_2) S_1 (p_2)\right]  \right\} \nonumber \\
&=& \frac12 (\delta_{p_1 k_1} \delta_{p_2 k_2 }+ \delta_{p_1 k_2} \delta_{p_2 k_1 } )   \left\{ 2  +\right. \nonumber \\
& & + 2 (S_1 (p_2) -1) + 2 (S_2 (p_2) -1) + 2 [S_1 (p_2 ) -1 ] [S_2
(p_2)-1 ]
\nonumber \\
& & + 2 (S_1 (p_1) -1) + 2 (S_2 (p_1) -1) + 2 [S_1 (p_1 ) -1 ] [S_2
(p_1)-1 ]
\nonumber \\
& &  + [S_1 (p_1) -1] [S_1 (p_2)-1] S_2 (p_2) + S_1 (p_1) [S_2 (p_1) -1] [S_2 (p_2)-1]   \nonumber \\
& &  + [ S_1 (p_1)-1] [S_2 (p_2)-1] + [S_1 (p_2)-1] S_2 (p_2) [S_2 (p_1)-1] S_1 (p_1)   \nonumber \\
& &  + [S_1 (p_2)-1] [S_1 (p_1)-1] S_2 (p_1) + S_1 (p_2) [S_2 (p_2)-1] [S_2 (p_1)-1]   \nonumber \\
& & \left. + [S_1 (p_2)-1] [S_2 (p_1)-1] + [ S_1 (p_1)-1] S_2 (p_1) [S_2 (p_2)-1] S_1 (p_2)  \right\} \nonumber \\
&=& (\delta_{p_1 k_1} \delta_{p_2 k_2 }+ \delta_{p_1 k_2}
\delta_{p_2 k_1 } )   S_1 (p_1) S_2 (p_1) S_2 (p_2) S_1 (p_2 ) =
S^{(1)}_{p_1 k_1} S^{(1)}_{p_2 k_2} + S^{(1)}_{p_1 k_2} S^{(1)}_{p_2
k_1}.
\label{s2red}
\eeq

Let us now prove that the two-photon scattering matrix on two atoms can be represented as a convolution of the two-photon scattering matrices on individual atoms, that is
\beq
S^{(2)}_{p_1 p_2 , k_1 k_2} & \stackrel{!}{=}& \frac{1}{2!} \int d k'_1 d k'_2 S^{(2)}_{2;p_1 p_2, k'_1 k'_2}  S_{1;k'_1 k'_2 , k_1 k_2}^{(2)} \nonumber \\
&=& \frac{1}{2!} \int d k'_1 d k'_2 \left( S_{2; p_1 k'_1}^{(1)} S_{2; p_2 k'_2}^{(1)} + S_{2; p_2 k'_1}^{(1)} S_{2; p_1 k'_2}^{(1)} + i \mathcal{T}^{(2)}_{2;p_1 p_2, k'_1 k'_2} \right) \nonumber \\
& & \times \left(S_{1; k'_1 k_1}^{(1)} S_{1;  k'_2 k_2}^{(1)} + S_{1; k'_1 k_2}^{(1)} S_{1; k'_2 k_1}^{(1)} +  i \mathcal{T}_{1;k'_1 k'_2 , k_1 k_2}^{(2)}  \right)\nonumber \\
&=& (\delta_{p_1 k_1} \delta_{p_2 k_2} + \delta_{p_1 k_2} \delta_{p_2 k_1}) S_{2} (p_1) S_{2} (p_2) S_{1} (p_1) S_{1} (p_2)  \nonumber \\
&+& S_2 (p_1) S_2 (p_2) i \mathcal{T}_{1; p_1 p_2, k_1 k_2} +S_1 (k_1) S_1 (k_2) i \mathcal{T}_{2; p_1 p_2, k_1 k_2} \nonumber \\
&-& \frac12  \int d k'_1 d k'_2 \mathcal{T}^{(2)}_{2;p_1 p_2, k'_1 k'_2} \mathcal{T}^{(2)}_{1; k'_1 k'_2 , k_1 k_2} .
\label{convol_twophot}
\eeq

The first term is obviously equal to the reducible contribution \eq{s2red}.
Let us now show that the second and the third terms together yield $S^{(2)irred}_{p_1 p_2, k_1 k_2}$. We get
\beq
& & 4 \delta_{p_1 +p_2 , k_1 +k_2} \left[  \frac{(2 \pi i g_1^2)^2}{4 \pi i} (E-2 \alpha_1) M_1 (p_1) M_1 (p_2) M_1 (k_1 ) M_1 (k_2 ) S_2 (p_1 ) S_2 (p_2) \right. \nonumber \\
& & \left. + \frac{(2 \pi ig_2^2)^2}{4 \pi i} (E-2 \alpha_2) M_2 (p_1) M_2 (p_2) M_2 (k_1 ) M_2 (k_2 ) S_1 (k_1 ) S_1 (k_2) \right] \nonumber \\
& & - \frac12 (4 \pi)^2 g_1^4 g_2^4 \delta_{p_1 +p_2 , k_1 + k_2} (E - 2 \alpha_1) (E - 2 \alpha_2 ) M_2 (p_1) M_2 (p_2)  M_1 (k_1) M_1 (k_2) \nonumber \\
& & \times \int d k'_1 M_2 (k'_1) M_2
(E - k'_1) M_1 (k'_1 ) M_1 (E - k'_1)  . 
\eeq
Evaluating the last integral we find
\beq
\int d k'_1 M_2 (k'_1) M_2 (E - k'_1) M_1 (k'_1 ) M_1 (E - k'_1) = -\frac{4 \pi i}{(E- \alpha_1 - \alpha_2) (E - 2 \alpha_1 ) (E - 2 \alpha_2)} , 
\eeq
and therefore
\beq
 & & 4 \delta_{p_1 +p_2 , k_1 +k_2} \left[  \frac{(2 \pi i g_1^2)^2}{4 \pi i} (E-2 \alpha_1) M_1 (p_1) M_1 (p_2) M_1 (k_1 ) M_1 (k_2 ) S_2 (p_1 ) S_2 (p_2) \right. \nonumber \\
& & + \frac{(2 \pi ig_2^2)^2}{4 \pi i} (E-2 \alpha_2) M_2 (p_1) M_2 (p_2) M_2 (k_1 ) M_2 (k_2 ) S_1 (k_1 ) S_1 (k_2)  \nonumber \\
& & -\left. \frac{  (2 \pi i g_1^2)^2  (2 \pi i g_2^2)^2}{2 \pi i} \delta_{p_1 +p_2 , k_1 + k_2}  \frac{M_2 (p_1) M_2 (p_2)  M_1 (k_1) M_1 (k_2)}{E- \alpha_1 - \alpha_2} \right] = S^{(2)irred}_{p_1 p_2, k_1 k_2}, 
\eeq
which reproduces the result contained in Eqs.~\eq{T2fin} and \eq{s2irred}.

Note  that the sequence of $S_1$ and $S_2$ in the convolution \eq{convol_twophot} corresponds
to the order in which the right-moving photons encounter the atoms $1$ and $2$ along the line of propagation. The presence of the irreducible
part $\mathcal{T}^{(2)}$ in the two-photon scattering makes the convolution
\eq{convol_twophot} noncommutative $S_2^{(2)} * S_1^{(2)} \neq S_1^{(2)} * S_2^{(2)}$, in contrast to the case of single-photon scattering \eq{convol_onephot}, where the scattering is commutative, $S_2^{(1)} * S_1^{(1)} = S_1^{(1)} * S_2^{(1)}$.

The result \eq{T2fin}-\eq{s2red} generalizes the corresponding expressions of Ref.~\cite{Y2} to arbitrary coupling strengths $g_1 \neq g_2$. Being equipped with the convolution property, one can write down as well an expression for the two-photon scattering on $M$ emitters with different $g_i$'s.

The calculations outlined above allows us to make the following general statement: the two-particle scattering matrix on an array of distributed emitters is a momentum space convolution of scattering matrices on individual emitters. This property has a deep reason: the  absence of backscattering for unidirectionally propagating photons. Therefore, it is expected to hold also in  arbitrary $N$-photon sector of scattering.

\section{Coherent light scattering}
In this section we consider scattering of coherent light off a single two-level emitter. The aim of this calculation is  twofold: at first, we describe the certain physical situation relevant for a discussion of the resonant fluorescence \cite{Mollow} in nanostructures, and, at second, we provide a generating functional for a $N$-particle scattering matrix.

We start from introducing operators of photons propagating inside a waveguide of a finite length $L$
\be
c_k = \frac{1}{\sqrt{2 \pi}} \int_{-L/2}^{L/2} d x a (x ) e^{- i k x}, \quad c_k^{\dagger} = \frac{1}{\sqrt{2 \pi}} \int_{-L/2}^{L/2} d x a^{\dagger} (x ) e^{i k x} ,
\label{op_c}
\ee
where $a (x)$ and $a^{\dagger} (x)$ are the Fourier transforms of the operators
$a_k$ and $a_k^{\dagger}$
\be
a (x) = \frac{1}{\sqrt{2 \pi}} \int d k a_k e^{i k x}, \quad a^{\dagger} (x)
= \frac{1}{\sqrt{2 \pi}} \int d k a_k^{\dagger} e^{-i k x} .
\label{ft}
\ee
The construction \eq{op_c} explicitly containing the finite length $L$ is very useful as it allows for a treatment of
different limits such as a finite photon number limit ($\bar{N} \equiv |\alpha_k|^2$=const, $L \to \infty$) and a finite photon density limit ($\bar{N}/L$=const at $\bar{N},L \to \infty$).

The operators \eq{op_c} obey the commutation relations
\beq
\left[ c_{k}, c_{p}^{\dagger} \right] &=& \left[ c_{k}, a_{p}^{\dagger} \right] = \left[ a_{p}, c_{k}^{\dagger} \right] =   \frac{1}{2 \pi i} \frac{e ^{i (k - p) L/2} - e^{-i (k - p) L/2}}{k - p} \equiv \delta_{\Delta} (k - p) , \\
\left[ a (x) , c_k^{\dagger} \right] &=& \frac{e^{i k x}}{\sqrt{2 \pi}} \Theta (-L/2 < x < L/2). \label{x_check2}
\eeq
where $\delta_{\Delta} (k - p)$ is a delta function broadened by $\Delta \equiv 2 \pi/L$ and reaching the peak value $1/\Delta$ at $k=p$. An expression of  the operator $c_k$ in terms of $a_k$ is given by 
\be
c_k = \int d p a_p \delta_{\Delta} (p-k).
\ee

As an initial state we choose a coherent state in the mode $k$
\be
| \alpha_k \rangle = e^{-|\alpha_k |^2/2} \sum_{n_k=0}^{\infty} \left( \frac{2 \pi}{L} \right)^{n_k/2} \frac{\alpha_k^{n_k} (c_k^{\dagger})^{n_k}}{n_k !} |0 \rangle = e^{-|\alpha_k |^2/2} \sum_{n_k=0}^{\infty} \frac{\alpha_k^{n_k} }{\sqrt{n_k !}} | n_k \rangle .
\label{init_coherent}
\ee
This state is normalized to unity, which is ensured by the finite waveguide's length $L$. It is a superposition of Fock states with different numbers of photons $n_k$, and therefore we introduce an operator $O$ which projects on-shell and accounts for the energy conservation in each photons' number sector. An outgoing state after the scattering off a two-level system is thus obtained by applying the operator $O$ followed by an action of the $T$-matrix, as prescribed by \eq{tmmN}, 
\beq
|\beta_k \rangle &=& (1 -2 \pi i T O)| \alpha_k \rangle  = |\alpha_k \rangle - 2\pi i e^{-|\alpha_k|^2/2} \sum_{n_k =1}^{\infty} \left( \frac{2 \pi}{L} \right)^{n_k/2} \alpha_k^{n_k} \sum_{n=1}^{n_k} g^{2n} \int \left( \prod_{i=1}^n d q_i d p_i  \right) \delta \left(\sum_{i=1}^n p_i - \sum_{i=1}^n q_i \right) \nonumber \\
&\times & a^{\dagger}_{p_1} \frac{1}{k n_k - H_b - \alpha} a_{q_1} \frac{1}{k n_k - H_b + i \eta} \ldots a_{p_n}^{\dagger} \frac{1}{k n_k - H_b - \alpha}  a_{q_n}  \frac{(c_k^{\dagger})^{n_k}}{n_k !}    | 0 \rangle  \\
&=& | \alpha_k \rangle - 2 \pi i e^{-|\alpha_k|^2/2} \sum_{n_k =1}^{\infty} \left( \frac{2 \pi}{L} \right)^{n_k/2} \alpha_k^{n_k} \sum_{n=1}^{n_k} \frac{(c_k^{\dagger})^{n_k-n}}{(n_k -n)!}  g^{2n}  \int \left( \prod_{i=1}^n d q_i d p_i  \right) \delta \left(\sum_{i=1}^n (p_i - q_i) \right)  \left( \prod_{i=1}^n \delta_{\Delta} (q_i - k) \right)
\nonumber \\
& \times & a^{\dagger}_{p_1} \frac{1}{[q_1 + \sum_{i=2}^n (q_i -p_i) - \alpha ] [ \sum_{i=2}^n (q_i - p_i) + i \eta]} a^{\dagger}_{p_2} \ldots a_{p_{n-1}}^{\dagger} \frac{1}{[q_{n-1} +q_n  - p_n - \alpha] [q_n  - p_n + i \eta ]}  a_{p_n}^{\dagger}   \frac{1}{q_n - \alpha}| 0 \rangle . \nonumber 
\eeq
Exchanging the order of summations $\sum_{n_k=1}^{\infty} \sum_{n=1}^{n_k} = \sum_{n=1}^{\infty} \sum_{n_k = n}^{\infty} $ and shifting $n_k \to n_k +n$ we find
\beq
| \beta_k \rangle &=& |\alpha_k \rangle -2 \pi i e^{-|\alpha_k|^2/2}  \sum_{n=1}^{\infty} \sum_{n_k =0}^{\infty} \left( \frac{2 \pi}{L} \right)^{(n_k +n)/2} \alpha_k^{n_k +n}  \frac{(c_k^{\dagger})^{n_k}}{n_k !}  g^{2n} \int \left( \prod_{i=1}^n d q_i d p_i  \right) \delta \left(\sum_{i=1}^n (p_i - q_i) \right)  \left( \prod_{i=1}^n \delta_{\Delta} (q_i - k) \right) \nonumber \\
& \times & a^{\dagger}_{p_1} \frac{1}{[q_1 + \sum_{i=2}^n (q_i - p_i) - \alpha ] [\sum_{i=2}^n (q_i - p_i ) + i \eta]} a^{\dagger}_{p_2} \ldots a_{p_{n-1}}^{\dagger} \frac{1}{[q_{n-1} + q_n  - p_n - \alpha] [q_n  - p_n + i \eta ]}  a_{p_n}^{\dagger} \frac{1}{q_n -\alpha}  | 0 \rangle \nonumber \\
&=& |\alpha_k \rangle - 2 \pi i \sum_{n=1}^{\infty} \left( \frac{2 \pi}{L} \right)^{n/2} \alpha_k^{n}  g^{2n} \int \left( \prod_{i=1}^n d q_i d p_i  \right)
\delta \left(\sum_{i=1}^n (p_i - q_i ) \right)  \left( \prod_{i=1}^n \delta_{\Delta} (q_i - k) \right)  \\
& \times & a^{\dagger}_{p_1} \frac{1}{[q_1 + \sum_{i=2}^n (q_i- p_i) - \alpha ] [\sum_{i=2}^n (q_i - p_i ) + i \eta]} a^{\dagger}_{p_2} \ldots a_{p_{n-1}}^{\dagger} \frac{1}{[q_{n-1}  + q_n  - p_n - \alpha] [q_n  - p_n + i \eta ]}  a_{p_n}^{\dagger}  \frac{1}{q_n - \alpha} | \alpha_k \rangle . \nonumber 
\eeq

Let us now consider the state
\beq
| \gamma_k^{(n)} \rangle &=& \int d p_1 \ldots d p_n \int d q_1 \ldots d q_n \delta \left(\sum_{i=1}^n (p_i - q_i) \right) \left( \prod_{i=1}^n \delta_{\Delta} (q_i - k) \right) \label{gam_st} \\
& \times & \frac{1}{[q_1 + \sum_{i=2}^n (q_i - p_i) - \alpha ] [\sum_{i=2}^n (q_i - p_i) + i \eta]} \ldots \frac{1}{[q_{n-1} +q_n  - p_n - \alpha] [q_n  - p_n + i \eta ]} \frac{1}{q_n - \alpha}
a_{p_1}^{\dagger} \ldots a_{p_n}^{\dagger} | \alpha_k \rangle.  \nonumber
\eeq
Performing the inverse Fourier transformation [cf. \eq{ft}]
\be
a_{p_i}^{\dagger} = \frac{1}{\sqrt{2 \pi}} \int d x_i a^{\dagger} (x_i ) e^{i p_i x_i},
\label{ift}
\ee
and substituting it into \eq{gam_st}, we obtain
\beq
| \gamma_k^{(n)} \rangle &=& \frac{1}{(2 \pi)^{n/2}} \int d x_1  \ldots  d x_n \int d p_1 \ldots d p_n \int d q _1 \ldots d q_n \delta \left(\sum_{i=1}^n (p_i - q_i )\right) \left( \prod_{i=1}^n \delta_{\Delta} (q_i - k) \right)
 \nonumber \\
& \times &  \frac{1}{[q_1 + \sum_{i=2}^n (q_i - p_i) - \alpha ] [\sum_{i=2}^n (q_i - p_i) + i \eta]} \ldots  \frac{1}{[q_{n-1} + q_n  - p_n - \alpha] [q_n  - p_n + i \eta ]}  \frac{1}{q_n - \alpha} 
\nonumber \\
& \times & e^{i p_1 x_1 + \ldots + i p_n x_n} a^{\dagger} (x_1 ) \ldots a^{\dagger} (x_n) | \alpha_k \rangle.
 \label{gam_st2}
\eeq
Introducing new variables $P_j = \sum_{i=j}^n p_i$ and $Q_j = \sum_{i=j}^n q_i$, 
such that  $p_i = P_i - P_{i+1}$ and $q_i =Q_i -Q_{i+1}$, we cast \eq{gam_st2} to
\beq
| \gamma_k^{(n)} \rangle &=& \frac{1}{(2 \pi)^{n/2}} \int d x_1  \ldots  d x_n \int d P_1 \ldots d P_n \int d q_1 \ldots d q_n \delta \left(P_1 - Q_1 \right) \left( \prod_{i=1}^n \delta_{\Delta} (q_i - k) \right)
 \nonumber \\
& \times & e^{i (P_1 -P_2) x_1} \frac{1}{[Q_1 - P_2 - \alpha ] [Q_2 -P_2 + i \eta]} \ldots  \frac{1}{[Q_{n-1}  - P_n - \alpha] [Q_n  - P_n + i \eta ]} e^{i P_n x_n} \frac{1}{Q_n -\alpha}
\nonumber \\
& \times & a^{\dagger} (x_1 ) \ldots a^{\dagger} (x_n) | \alpha_k \rangle \\
&=& \frac{1}{(2 \pi)^{n/2}} \int d x_1  \ldots d x_n \int d q_1 \ldots d q_n 
\left( \prod_{i=1}^n 
 \delta_{\Delta} (q_i - k) \right) e^{i Q_1 x_1} \frac{1}{Q_n - \alpha}\nonumber \\
& \times & \left( \int d P_2 \frac{e^{-i P_2 (x_1 -x_2)}}{[Q_1 - P_2 - \alpha ] [Q_2 -P_2 + i \eta]} \right) \ldots \left( \int d P_n \frac{e^{-i P_n (x_{n-1} -x_n)}}{[Q_{n-1} - P_n - \alpha ] [Q_n  -P_n + i \eta]}\right)  a^{\dagger} (x_1 ) \ldots a^{\dagger} (x_n) | \alpha_k \rangle . \nonumber 
\eeq

Each integral over $d P_j$
does not vanish, if one  can close the contour of integration in the upper half plane. It is only possible for $\Delta x_j \equiv x_{j}-x_{j-1} >0$. Then we get
\beq
\int d P_j \frac{e^{i P_j \Delta x_j}}{[Q_{j-1} - P_j - \alpha ] [Q_j  -P_j + i \eta]} = \Theta (x_j - x_{j-1})  2 \pi i e^{i Q_j \Delta x_j} \frac{e^{i (q_{j-1}-\alpha) \Delta x_j}-1}{q_{j-1} -\alpha},
\eeq
and therefore
\beq
| \beta_k \rangle &=& |\alpha_k \rangle + \sum_{n=1}^{\infty}  \left( - 2 \pi i g^2 \bar{\alpha}_k \right)^{n} \int d x_1  \ldots d x_n \int d q_1 \ldots d q_n \left( \prod_{i=1}^n \delta_{\Delta} (q_i - k) \right)
\Theta (x_n > \ldots >x_1)  \nonumber \\
& \times & e^{i q_1 x_1 + \ldots + i q_n x_n} \frac{1}{q_n - \alpha } \prod_{j=2}^n \frac{1- e^{i (q_{j-1} -\alpha) \Delta x_j}}{q_{j-1} -\alpha}
a^{\dagger} (x_1 ) \ldots a^{\dagger} (x_n) | \alpha_k \rangle \nonumber \\
&=& |\alpha_k \rangle + \sum_{n=1}^{\infty}  \left( - 2 \pi i g^2 \bar{\alpha}_k \right)^{n} \int d x_1  \ldots d x_n  \Theta (x_n > \ldots >x_1)  \nonumber \\
& \times &
\left( \int d q_n \frac{e^{i q_n x_n}}{q_n - \alpha} \delta_{\Delta} (q_n - k) \right) \left( \prod_{j=1}^{n-1} \int d q_j e^{i q_j x_j} \frac{1- e^{i (q_{j} -\alpha) \Delta x_{j+1}}}{q_{j} -\alpha} \delta_{\Delta} (q_j - k) \right)
a^{\dagger} (x_1 ) \ldots a^{\dagger} (x_n) | \alpha_k \rangle ,
\label{beta_k1}
\eeq
where $\bar{\alpha}_k = \alpha_k/\sqrt{L}$.

It remains to evaluate integrals
\beq
& & \int d q_j \frac{e^{i q_j x_j}}{q_j - \alpha} \delta_{\Delta} (q_j - k) = \frac{e^{i k x_j}}{4 \pi i} \int d q_j \frac{e^{i q_j (x_j+L/2)} - e^{-i q_j (L/2 - x_j)} }{q_j + k - \alpha} \left[ \frac{1}{q_j +i 0^+} +\frac{1}{q_j - i 0^+} \right] \nonumber \\
&=& \frac12 e^{i k x_j} \left[ \frac{1}{k-\alpha} \Theta (x_j +L/2) + \left( 
\frac{2}{k - \alpha} e^{- i (k-\alpha) (x_j + L/2)}- \frac{1}{k - \alpha} \right) \Theta (-x_j - L/2) \right. \nonumber \\
& & \left. \qquad \quad + \left( - \frac{2}{k - \alpha} e^{i (k-\alpha) (L/2 - x_j)} + \frac{1}{k - \alpha}\right) \Theta (L/2 - x_j) - \frac{1}{k-\alpha} \Theta (x_j - L/2)\right] \nonumber \\
&=& \frac{e^{i k x_j}}{k -\alpha} \left[ 1 - e^{i (k-\alpha) (L/2-x_j)}\right] \Theta (-L/2 < x_j <L/2) \nonumber \\
&+& \frac{e^{i k x_j}}{k -\alpha} \left[ e^{-i (k-\alpha) (x_j + L/2)} - e^{i (k-\alpha) (L/2-x_j)}\right] \Theta (x_j < -L/2)
\label{int_no1}
\eeq
and
\beq
& & - \int d q_j \frac{e^{i q_j x_j} e^{i (q_j - \alpha) \Delta x_{j+1}}}{q_j - \alpha} \delta_{\Delta} (q_j - k) = - e^{- i \alpha \Delta x_{j+1}} \int  d q_j \frac{e^{i q_j x_{j+1}}}{q_j - \alpha} \delta_{\Delta} (q_j - k)
\nonumber \\
&=& - e^{- i \alpha \Delta x_{j+1}} \frac{e^{i k x_{j+1}}}{k -\alpha} \left[ 1 - e^{i (k-\alpha) (L/2-x_{j+1})}\right] \Theta (-L/2 < x_{j+1} <L/2)
\nonumber \\
& & - e^{- i \alpha \Delta x_{j+1}} \frac{e^{i k x_{j+1}}}{k -\alpha} \left[ e^{-i (k-\alpha) (x_{j+1} + L/2)} - e^{i (k-\alpha) (L/2-x_{j+1})}\right] \Theta (x_{j+1} < -L/2).
\label{int_no2}
\eeq
Summing up the both terms \eq{int_no1} and \eq{int_no2} yields
\beq
& & \Theta (x_{j+1} - x_j) \int d q_j \frac{e^{i q_j x_j} [1 - e^{i (q_j - \alpha) \Delta x_{j+1}}]}{q_j - \alpha} \delta_{\Delta} (q_j - k) \nonumber \\
&=& \Theta (-L/2 < x_j < x_{j+1} < L/2) \frac{e^{i k x_j}}{k - \alpha} \left[1 - e^{i (k-\alpha) \Delta x_{j+1}} \right] \nonumber \\
&+& \Theta (x_j < - L/2 < x_{j+1} < L/2) \frac{e^{i k x_j}}{k - \alpha}
\left[ e^{-i (k - \alpha ) (x_j +L/2)} - e^{i (k-\alpha) \Delta x_{j+1}}\right].
\label{int_no3}
\eeq
Note that there is no contribution to \eq{int_no3} when both $x_j$ and $x_{j+1}$ are smaller than $-L/2$.

Thus, we finally obtain an expression for the state $| \beta_k \rangle$ which emerges after scattering of the initially prepared coherent state \eq{init_coherent} off the two-level system
\beq
| \beta_k \rangle &=& | \alpha_k \rangle + \sum_{n=1}^{\infty} \left( - \frac{2 \pi i g^2 \bar{\alpha}_k}{k-\alpha}\right)^n \int d x_1 \ldots d x_n e^{i k (x_1 + \ldots +x_n)} \nonumber \\
&\times & \left[ \Theta
(L/2 > x_n > \ldots > x_2 >x_1 > - L/2) \prod_{j=1}^n \left( 1 - e^{i (k-\alpha) \Delta x_{j+1}}\right) \right. \nonumber \\
& & \left. + \Theta
(L/2 > x_n > \ldots > x_2 > - L/2 >x_1)  e^{- i (k-\alpha) (x_1 +L/2)} \left( 1- e^{i (k-\alpha) (x_2 +L/2)} \right) \prod_{j=2}^n \left( 1 - e^{i (k-\alpha) \Delta x_{j+1}}\right)\right] \nonumber \\
& \times & a^{\dagger} (x_1 ) \ldots a^{\dagger} (x_n) | \alpha_k \rangle,
\label{fin_coherent}
\eeq
where $x_{n+1} \equiv L/2$.

The state \eq{fin_coherent} has a remarkable property: when two coordinates in the integrand approach each other, it vanishes. This property lies in the origin of antibunching of photons which is conventionally observed in the density-density correlation functions \cite{Mollow},\cite{Yu-preprint}. An occurrence of the two contributions with $x_1 > -L/2$ and $x_1 <-L/2$ has been previously remarked in \cite{Yu-preprint}, both being important for a proper normalization
of the state \eq{fin_coherent}.

The explicit expression for the outgoing state \eq{fin_coherent} opens a possibility for a study of correlation functions as well as  photons' statistics in all parametric regimes, which will be a subject of a subsequent publication \cite{PRYG}.

\section{Conclusions}
We have developed the scattering approach to problems of propagating bosons in one dimensional geometry. We have derived the general form of the $T$-matrix (non-trivial part of the scattering matrix), Eq.~(\ref{eq:main-formula}), under the following assumptions about the spectrum of bosons: the spectrum is linear, chiral, and infinite. Our formalism is complimentary to the Bethe ansatz solutions \cite{RuYu} and to the traditional approaches based on the equations of motion. 

We have applied the developed formalism to several specific examples including emitters with two- and three-level structures. The emitters can be either distributed across the 1D channel or concentrated in a tiny region of space. We have shown that the scattering results in projecting the state of the emitter onto the specific -- dark -- state which does not emit. 

We have also shown that the one- and two-particle scattering matrices off  two emitters can be represented as a convolution of scattering matrices corresponding to individual emitters. Hereby the microscopic properties of different emitters can vary (i.e. coupling constants to photons, detunings, level structure). We conjecture that this property generically holds for multi-emitter arrays as well as for the multiparticle scattering with $N>2$ photons. We are going to elaborate more on these issues in future studies \cite{PRYG}.

The developed approach can be applied to the arbitrary initial state, which can either conserve the particle number or not. In the case of number-conserving initial state, we observe  the formation of photonic bound states, which is reflected in the emergence of a pole of the $S$-matrix involving in its argument an energy of more than one individual photon. In the case of the coherent light scattering, we clearly observe the  fermionized behavior typical to the Tonks-Girardeau gas \cite{TG} discussed in the photonic context in \cite{NP}. 

\section{Acknowledgements}
We are grateful to Matous Ringel and Vladimir Yudson for fruitful discussions. MP acknowledges the financial support from DFG-FG 723. VG is supported by the Swiss National Science Foundation.

\end{document}